\documentclass[12pt]{article}
\pdfoutput=1
\usepackage[]{hyperref}
\usepackage{xcolor}
\usepackage[]{putex}
\usepackage[utf8]{inputenc}
\usepackage{subcaption}
\makeatletter
\DeclareRobustCommand*{\bfseries}{%
  \not@math@alphabet\bfseries\mathbf
  \fontseries\bfdefault\selectfont
  \boldmath
}
\makeatother
\hypersetup{
  colorlinks=true,
  linkcolor=blue,
  citecolor=green!60!black,
  urlcolor=cyan!70!black,
  linktoc=all
  }

\numberwithin{equation}{section}

\input{glyphtounicode}
\begin{document}

\preprint{PUPT-2565}

\title{$p$-adic Mellin Amplitudes}
\authors{Christian Baadsgaard Jepsen$^1$\footnote{\tt cjepsen@princeton.edu} $\&$ Sarthak Parikh$^2$\footnote{\tt sparikh@caltech.edu}}
\institution{PU}{$^1$Joseph Henry Laboratories, Princeton University, Princeton, NJ 08544, USA}
\institution{Caltech}{$^2$Division of Physics, Mathematics and Astronomy,\cr\hskip0.06in California Institute of Technology, Pasadena, CA 91125, USA}

\abstract{In this paper, we propose a $p$-adic analog of Mellin amplitudes for scalar  operators, and present the computation of the general contact amplitude as well as arbitrary-point tree-level amplitudes for bulk diagrams involving up to three internal lines, and along the way obtain the $p$-adic version of the split representation formula. These amplitudes share noteworthy  similarities with the usual (real) Mellin amplitudes for scalars, but are also significantly simpler, admitting closed-form expressions where none are available over the reals. The dramatic simplicity can be attributed to the absence of descendant fields in the $p$-adic formulation.}

\date{\today}

\maketitle

{\hypersetup{linkcolor=black}
\tableofcontents
}

\section{Introduction and Summary}
\label{INTRODUCTION}

Anti-de Sitter/conformal field theory (AdS/CFT) duality~\cite{Maldacena:1997re,Gubser:1998bc,Witten:1998qj,Aharony:1999ti} provides a powerful framework for investigating the properties of correlators, the basic observables, in strongly coupled CFTs. 
Early work in the subject~\cite{Muck:1998rr,Liu:1998ty,Freedman:1998bj,DHoker:1998bqu,Liu:1998th,DHoker:1998ecp,DHoker:1999bve,DHoker:1999kzh,DHoker:1999mqo,DHoker:1999mic,Arutyunov:2002fh} laid the foundation for computational techniques, especially in the context of the holographic evaluation of correlators via bulk Feynman diagram methods. 
Traditionally, CFT correlators are obtained in position space, which though physically intuitive, often falls short of utilizing the full power of conformal symmetry. Consequently, despite major advances in evaluating holographic correlators in position space, the study and computation of arbitrarily complicated bulk diagrams remained a challenging task. 
But beginning with the work of Mack~\cite{Mack:2009mi}, developed further in Refs.~\cite{Penedones:2010ue,Fitzpatrick:2011ia,Paulos:2011ie,Fitzpatrick:2011hu,Nandan:2011wc,Fitzpatrick:2011dm,Fitzpatrick:2012cg,Costa:2012cb,Goncalves:2014rfa} in the holographic context, Mellin amplitudes emerged as an effective tool in this regard. Analogous to momentum space for flat space scattering amplitudes, Mellin space can be regarded as the natural space for studying scattering amplitudes in AdS, one reason being that it manifestly takes into account the conformal symmetry of the underlying theory. While position space correlators are written as functions of conformally invariant cross-ratios constructed out of the boundary insertion points $x_i$, the Mellin amplitude ${\cal M}$ depends on Mandelstam-like invariants defined in terms of the Mellin variables $\gamma_{ij}$ -- indeed, the number of conformally invariant cross-ratios in position space matches the number of independent Mandelstam-like variables in Mellin space. An $\mathcal{N}$-point position space correlator ${\cal A}(\{x_i\})$ is represented as the inverse Mellin transform of the Mellin amplitude ${\cal M}(\{\gamma_{ij}\})$, defined (schematically) via the contour integral
\eqn{MellinAmpDef}{
\mathcal{A}(\{x_i\})=\int[d\gamma]\,\mathcal{M}(\{\gamma_{ij}\})\prod_{1\leq i < j \leq {\cal N}}\frac{\Gamma(\gamma_{ij})}{|x_i-x_j|^{2\gamma_{ij}}}\,, 
}
where the measure $[d\gamma]$ is over the Mellin variables $\gamma_{ij}$, which are integrated along contours parallel to the imaginary axis according to a well-defined prescription.  
In the early papers~\cite{Fitzpatrick:2011ia,Paulos:2011ie}, a set of ``Feynman rules'' were derived which yield, in principle, the Mellin amplitude for any bulk-diagram at tree-level, and to this date the study of Mellin space has continued to yield new insights into the structure of correlators and holography, see e.g.\ Refs.~\cite{Gopakumar:2016cpb,Aharony:2016dwx,Alday:2017gde,Alday:2017xua,Rastelli:2016nze,Rastelli:2017udc,Faller:2017hyt,Chen:2017xdz,Yuan:2017vgp,Yuan:2018qva}.

Recently,
the framework of holography was extended to the so-called $p$-adic AdS/CFT correspondence~\cite{Gubser:2016guj,Heydeman:2016ldy,Gubser:2017tsi}. 
In the simplest setting, the classical bulk geometry is described by the Bruhat--Tits tree, essentially an infinite $(p+1)$-regular graph without any loops, in the place of vacuum (Euclidean) AdS.\footnote{We refer the reader to Ref.~\cite{Gubser:2016guj} for a discussion on how this tree structure emerges as a course-graining at AdS length scales of a continuum $p$-adic bulk (see also Ref.~\cite{Heydeman:2016ldy}). 
For a description of the bulk in other non-trivial geometries such as black-hole backgrounds, see Refs.~\cite{Manin:2002hn,Heydeman:2016ldy,Gubser:2016htz}.} The projective line over $p$-adic numbers, in place of reals, is interpreted as the boundary of the tree.\footnote{Arguably, this version of $p$-adic holography is similar in certain aspects, such as the structure of the global conformal group, to AdS$_3$/CFT$_2$ or even AdS$_2$/CFT$_1$. However, one can make contact with certain aspects of higher dimensional AdS$_{n+1}$/CFT$_n$ holography for any $n$, if one considers the degree $n$ (unramified) extension of $p$-adic numbers on the boundary, corresponding to a bulk given by the Bruhat--Tits tree associated with the unramified extension~\cite{Gubser:2016guj,Gubser:2017tsi} (see also Ref.~\cite{Gubser:2017vgc} for a non-trivial example of an interacting $p$-adic CFT defined on such a boundary).} 
Just like in the usual AdS/CFT prescription, boundary correlators may be obtained via holographic computations. Surprisingly, the position space correlators in the $p$-adic formulation are strikingly similar to their real analogs, not just with respect to the kinematics (such as the functional dependence on coordinates) but also with respect to the dynamics (such as the functional form of the OPE coefficients)~\cite{Gubser:2016guj,Gubser:2017tsi,Gubser:2017vgc}.\footnote{In fact, one may be tempted to develop a dictionary to translate results back and forth between the two formulations, reminiscent of related observations made earlier in the context of $p$-adic string theory~\cite{Freund:1987kt,Freund:1987ck,Brekke:1988dg,Brekke:1988yb,Brekke:1993gf}.}
At the same time, the $p$-adic results are much simpler, so that for instance  closed-form expressions are usually available for position space correlators, in stark contrast with the situation in real AdS/CFT.
Thus, in certain respects, the $p$-adic formulation provides a simpler, computationally efficient window into the usual formulation of holography over the reals.

\vspace{1em}
Given the important role Mellin amplitudes have played in the usual AdS/CFT correspondence and the similarities between the position space correlators in the $p$-adic and real formulations of holography, it is natural to ask whether $p$-adic versions of Mellin space and Mellin amplitudes exist and whether they can prove as fruitful in the context of $p$-adic AdS/CFT.
The primary goal of this paper is to  develop the framework of $p$-adic Mellin amplitudes,\footnote{We should emphasize that what we refer to as the $p$-adic Mellin amplitude is fundamentally different from what Ref.~\cite{Dutta:2017bja} denotes by the same name. The Mellin variables in our formalism live on a different manifold, and the pole structure of the Mellin amplitudes we derive also differs entirely from the one mentioned in Ref.~\cite{Dutta:2017bja}.} and to demonstrate the
similarities that the $p$-adic and real Mellin amplitudes share with each other. 
In the remainder of this section, we begin by motivating and proposing the definition of $p$-adic Mellin amplitudes (in section \ref{MELLINSPACE}), then proceed to recalling  the main properties of $p$-adic numbers and the correlators of $p$-adic AdS/CFT which will be needed (in section \ref{p-adicRecap}), before finally providing a summary of the main results of this paper (in section \ref{SUMMARY}).

\subsection{Mellin space and local zeta functions}
\label{MELLINSPACE}

In the standard AdS$_{n+1}$/CFT$_n$ formulation, to any $\mathcal{N}$-point position space amplitude $\mathcal{A}(\{x_i\})$ there corresponds a Mellin amplitude $\mathcal{M}$, which is a function of complex Mellin variables $\gamma_{ij}$, with indices $i$ and $j$ running from 1 to $\mathcal{N}$. The Mellin variables $\gamma_{ij}$  satisfy the constraints
\eqn{MellinVarConstraints}{
\gamma_{ij}=\gamma_{ji}\,, \qquad   \sum_{j=1}^\mathcal{N}\gamma_{ij}=0 \quad {\rm and }\quad  \gamma_{ii}=-\Delta_i \quad {\rm (no\ sum\ over\ } i) \qquad i=1,\ldots, \mathcal{N}\,. 
}
If the bulk space-time dimension is $(n+1)$, then provided $n$ is sufficiently large, the conditions in \eno{MellinVarConstraints} admit $\mathcal{N}(\mathcal{N}-3)/2$ independent Mellin variables, which is the number of independent conformally invariant cross-ratios constructed out of $\mathcal{N}$ points.\footnote{More precisely, we assume $n+1 \geq {\cal N}$, 
otherwise there are $n\mathcal{N}-\frac{1}{2}(n+1)(n+2)$ conformally invariant cross-ratios (see, e.g.\ Ref.~\cite{Rastelli:2017udc}).}
The standard trick for solving the constraints is to introduce fictitious $(n+1)$-dimensional momenta $k_i$ (where we have suppressed the space-time Lorentz index) such that
 \eqn{momConstraints}{
 k_i\cdot k_j = \gamma_{ij} \qquad \sum_{i=1}^\mathcal{N} k_i = 0 \qquad i,j \in \{1,\ldots,\mathcal{N}\}\,,
 }
which supplemented with \eno{MellinVarConstraints}, implies the ``on-shell condition''
 \eqn{OnshellMom}{
k_i^2 = k_i \cdot k_i = -\Delta_i \qquad i\in \{1,\ldots,\mathcal{N}\}\,.
 }
 For $n+1 \geq {\cal N}$, the number of independent momentum degrees of freedom, which is the same as the number of independent Mandelstam invariants constructed out of momenta from the set $\{ k_i : i \in \{1,\ldots,\mathcal{N}\} \}$, is precisely $\mathcal{N}(\mathcal{N}-3)/2$.
Such Mandelstam invariants $s_{i_1 \ldots i_K}$, associated with a subset $S = \{i_1,\ldots, i_K\} \subseteq \{1,...,\mathcal{N}\}$ are defined to be
\eqn{MandelstamDef}
{
s_{i_1 \ldots i_K} \equiv -\left(\sum_{i\in S} k_i\right)^2 = \sum_{i\in S}\Delta_i - 2\sum_{\substack{i,j \in S,\\ i<j}}\gamma_{ij}\,.
}
We note that analogously to flat space scattering amplitudes, Mellin amplitudes exhibit dependence on Mellin variables $\gamma_{ij}$ only via such Mandelstam invariants.

As indicated previously, the Mellin space amplitudes ${\cal M}$  (over the reals) are defined via \eno{MellinAmpDef}, repeated below for convenience
\eqn{ArcMellin}
{
\mathcal{A}(\{x_i\})=\int[d\gamma]\,\mathcal{M}(\{\gamma_{ij}\})\prod_{1\leq i < j \leq \mathcal{N}}\frac{\Gamma(\gamma_{ij})}{|x_i-x_j|^{2\gamma_{ij}}}\,,
}
where
\eqn{MellinMeasure}{
[d\gamma] \equiv \prod_{(ij)}^{\mathcal{N}(\mathcal{N}-3)/2} {d\gamma_{ij} \over 2\pi i}
}
denotes a $\frac{\mathcal{N}(\mathcal{N}-3)}{2}$-dimensional measure over the independent Mellin variables $\gamma_{ij}$, and the individual contours are chosen to lie parallel to the imaginary axis, such that they separate out the semi-infinite sequences of poles arising from the Euler gamma functions in \eno{ArcMellin}. In Euclidean signature, the coordinate dependent factor in \eno{ArcMellin}, $|x_i - x_j|^2 = (x_i - x_j) \cdot (x_i - x_j)$ denotes the $L^2$-norm squared of the vector $x_i-x_j \in \mathbb{R}^n$.

It is convenient to factor out the product of Euler gamma functions $\Gamma(\gamma_{ij})$ from the definition of the Mellin amplitude ${\cal M}$  as shown in \eno{ArcMellin}. The gamma function $\Gamma(\gamma_{ij})$ in \eno{ArcMellin} has simple poles at $\gamma_{ij}=0,-1,-2,\ldots$ in the complex plane. 
Evaluating the contour integrals in \eno{ArcMellin}, it turns out the residues at the poles of these gamma functions generate, for large $N$ CFTs, precisely the double-trace contribution to the correlator in position space.
Consequently the Mellin amplitude is restricted to the single-trace sector, with poles of the amplitude corresponding precisely to the exchange of single-trace operators and their descendants in the intermediate channels.
As a result, the Mellin amplitude of an arbitrary-point contact diagram between scalar primaries is simply a constant, i.e.\ independent of Mellin variables $\gamma_{ij}$; in contrast, in position space, already the four-point contact diagram is represented by appropriate $D$-functions.

One of the peculiar features of $p$-adic AdS/CFT correspondence is that in a $p$-adic CFT,\footnote{By $p$-adic CFTs, we mean CFTs where the fundamental fields and operators are maps ${\cal O} : V \to \mathbb{R}$, where $V=\mathbb{Q}_p$ or some field-extension of $\mathbb{Q}_p$~\cite{Melzer:1988he}.} the OPE of two (scalar) operators features neither descendants nor multi-trace primaries  containing any derivatives~\cite{Melzer:1988he,Gubser:2017tsi}. 
Consequently, the conformal block decomposition at leading order in $1/N$ obtains contributions (aside from single-trace operators) only from double-trace operators of the form, ${\cal O}_A {\cal O}_B$, i.e.\ all derivatives are absent. 
Mathematically, this can be attributed to the observation that in $p$-adic AdS/CFT, the role of the Euler gamma function is played by the so-called ``local zeta function at a finite place'' (which we will refer to as the $p$-adic local zeta function)~\cite{Gubser:2016guj,Gubser:2017vgc,Gubser:2017tsi}
\eqn{zetap}
{
\zeta_p(z) \equiv \frac{1}{1-p^{-z}} \qquad z \in \mathbb{C},
}
where $p$ is a fixed prime number (denoting the ``finite place'' of the local zeta function), and the fact that it has a single simple pole along the real axis, at $z=0$.

A product over all the finite places $p$ of the local zeta function gives (via the Euler product formula) the Riemann zeta function,
\eqn{RiemannZeta}{
\zeta(z) = \sum_{n=1}^\infty {1 \over n^z} = \prod_{p {\rm \ prime}} \zeta_p(z)\,,
}
which has a simple pole at $z=1$.
The infinite sum in \eno{RiemannZeta} converges for $\Re(z)>1$, and then $\zeta(z)$ is extended to the entire complex plane via meromorphic continuation.
The Euler gamma function $\Gamma$, and the local zeta functions $\zeta_p$ can be combined together to define the ``completed zeta function'' (also referred to as the ``adelic zeta function'') via\footnote{Sometimes the completed zeta function $\zeta_{\mathbb{A}}$ is denoted $\zeta^*$ in the literature.}
\eqn{AdelicZeta}{
\zeta_{{\mathbb A}}(z) \equiv  \pi^{-z/2}\,\Gamma\left(\frac{z}{2}\right) \zeta(z) = \zeta_\infty(z) \prod_p \zeta_p(z)\,,
}
which satisfies the functional equation
\eqn{FE}{
\zeta_{{\mathbb A}}(z) = \zeta_{{\mathbb A}}(1-z)\,.
}
In \eno{AdelicZeta} we have defined the ``local zeta function at infinity'', as follows
\eqn{zetainfty}{
\zeta_\infty(z) \equiv \pi^{-z/2}\,\Gamma\left(\frac{z}{2}\right).
}
It is clear from \eno{AdelicZeta} that the completed zeta function treats the Euler gamma function $\Gamma(z/2)$ on the same footing as each of the local zeta functions at finite places, $\zeta_p(z)$.\footnote{The tree-level $N$-tachyon amplitudes in ($p$-adic) open string theory~\cite{Freund:1987kt,Freund:1987ck,Brekke:1988dg,Zabrodin:1988ep,Brekke:1993gf} can be expressed entirely in terms of the local zeta functions described here, and in fact the functional equation \eno{FE} plays an important role in the context of adelic strings~\cite{Freund:1987ck,Brekke:1988yb}, as it is central to the simple product rule satisfied by the channel symmetric Veneziano amplitude ~\cite{Freund:1987ck}: $A_\infty^{(4)}(k_i) \prod_p A_p^{(4)}(k_i) = 1$, where $A_\infty^{(4)}$ is the ordinary channel-symmetric Veneziano amplitude and $A_p^{(4)}$ is the corresponding Veneziano amplitude in $p$-adic string theory.}

It was observed in Refs.~\cite{Gubser:2016guj,Gubser:2017tsi,Gubser:2017vgc} that the structure constants and anomalous dimensions in the conformal block decomposition of scalar correlators in the standard formulation of AdS/CFT over the reals may be repackaged in terms of $\zeta_\infty$ functions (this in turn essentially removes all awkward factors of $\pi$ appearing in various formulae), and analogously the same scalar correlators are expressed in terms of $\zeta_p$ functions in $p$-adic AdS/CFT. 
Curiously, one can essentially go back and forth between the two cases by switching $\zeta_\infty$ and $\zeta_p$ in the results (modulo some important details which we gloss over here; see Refs.~\cite{Gubser:2016guj,Gubser:2017tsi,Gubser:2017vgc} for details).\footnote{There are indications~\cite{Gubser:2017qed} (see also Refs.~\cite{Zabrodin:1988ep, Ruelle:1989dg,Marshakov:1989jz}) that the coefficients of fermionic correlators may, analogous to the scalar case, be expressed in terms of local factors associated with the Dirichlet $L$-function (i.e.\ the ``local Dirichlet $L$-functions at finite $p$'' and the ``local Dirichlet $L$-function at infinity''). The Dirichlet $L$-function is the simplest generalization of the Riemann zeta function, and it generalizes the infinite sum in \eno{RiemannZeta} by weighting each term in the series by a simple non-trivial multiplicative character  (see e.g.\ Ref.~\cite{Brekke:1993gf} for a simple introduction to the Dirichlet $L$-function). The Riemann zeta function corresponds to the choice of the trivial multiplicative character as the weight factor.}

These considerations suggest a natural candidate for the definition of $p$-adic Mellin amplitudes (which we will also denote by the symbol ${\cal M}$; it should be clear from the context whether we are referring to real or $p$-adic Mellin amplitudes):
\eqn{pMel}
{
\mathcal{A}(\{x_i\})=\int[d\gamma]\,\mathcal{M}(\{\gamma_{ij}\})\prod_{1\leq i < j \leq \mathcal{N}}\frac{\zeta_p(2\gamma_{ij})}{|x_i-x_j|_p^{2\gamma_{ij}}}\,,
}
where ${\cal A}$ is the position space correlator in  $p$-adic AdS/CFT. Note that the position- and Mellin-space amplitudes in $p$-adic AdS/CFT are by construction \emph{real-} and \emph{complex-valued} functions  (for $p$-adic valued coordinates $x_i$) respectively, just as in real AdS/CFT; we refer to them simply as $p$-adic amplitudes to distinguish them from the corresponding amplitudes in the usual formulation of AdS/CFT over the reals. The measure $[d\gamma]$ in \eno{pMel} is given by
\eqn{pMellinMeasure}{
[d\gamma] \equiv \prod_{(ij)}^{\mathcal{N}(\mathcal{N}-3)/2} {d\gamma_{ij} \over 2\pi i/(2\log p)}\,,
}
where the factor of $(2\log p)$ has been introduced for later convenience, and the integral in \eno{pMel} is still over $\frac{\mathcal{N}(\mathcal{N}-3)}{2}$ independent Mellin variables $\gamma_{ij}$ which satisfy \eno{MellinVarConstraints}. Compared to \eno{ArcMellin}, in \eno{pMel} we have essentially replaced the Euler gamma function $\Gamma(s)$ with $\zeta_p(2s)$, the $p$-adic local zeta function with twice the argument of the Euler gamma function in line with \eno{AdelicZeta}-\eno{zetainfty}, and replaced the $L^2$-norm $|\cdot|$ over the reals with the $p$-adic norm $|\cdot|_p$. The $p$-adic norm will be described in the next subsection.

 Importantly, we should point out that  the contour prescription in \eno{pMel} is somewhat different from the one described below \eno{ArcMellin}; it is convenient to let the Mellin variables live on a manifold different from $\mathbb{C}$. To see which manifold, consider the periodicity of the local zeta function $\zeta_p$. From its definition \eqref{zetap}, it is clear that $\zeta_p(z)$ is periodic in the imaginary direction with periodicity ${2\pi}/{\log p}$, i.e.\
\eqn{}
{
\zeta_p\left(z+i\frac{2\pi}{\log p}\right)=\zeta_p(z).
}
As a consequence, it will be convenient to identify Mellin variables $\gamma_{ij}$ up to the addition of integral multiples of $i \pi/\log p$.  Thus we may choose the ``fundamental domain'' of $\gamma_{ij}$ to be $\mathbb{R} \times \left[-{\pi\over 2\log p}, {\pi \over 2\log p}\right)$. In other words, due to the periodic identification, we postulate: 
\begin{quote}
\emph{The Mellin variables for $p$-adic Mellin amplitudes live, not on the complex plane, but on an infinitely long horizontal cylinder, with circumference ${\pi}/{\log p}$}.
\end{quote}
The integration contours in \eqref{pMel} then turn out to be
circular contours winding once around the complex cylinder. On the fundamental domain, this corresponds to integration contours parallel to the imaginary axis, with the lower and upper limits of the imaginary part given by $-\frac{i\pi}{2\log p}$ and $\frac{i\pi}{2\log p}$, respectively. 
(Over the reals, the ``fundamental domain'' is the entire complex plane, and thus the contours run parallel to the imaginary axis from $-i\infty$ to $i\infty$, which curiously corresponds to taking the $p \to 1$ limit in the $p$-adic formulation.\footnote{For discussions on the $p\to 1$ limit in the context of $p$-adic string theory and $p$-adic AdS/CFT, see e.g.\ Refs.~\cite{Ghoshal:2006te,Bocardo-Gaspar:2017atv,Gubser:2016guj}.})
Just like in \eno{ArcMellin}, the contours are placed so that they separate out poles arising from different factors of the local zeta functions. 
This point is explained in detail via an explicit example in section \ref{EXAMPLE4PT}.
However, unlike the Euler gamma function which has a semi-infinite sequence of poles along the real axis, the $p$-adic local zeta function $\zeta_p(z)$ has only one (simple) pole at $z=0$ in the fundamental domain. This simplicity in the pole structure of the local zeta function $\zeta_p$ leads to great simplifications in the computations to follow, and accords the $p$-adic formulation of Mellin amplitudes its remarkable computational power. 

Before closing this subsection, we point out one more motivation for taking the complex Mellin variables to live on a cylindrical manifold in the case of $p$-adic Mellin amplitudes. The $p$-adic versions of the two (real) Barnes lemmas, which in the real case provide formulae for contour integrals over products of Euler gamma functions on the complex plane, take essentially the same form as their real analogs once we replace the Euler gamma functions with the appropriate local zeta functions $\zeta_p$, as long as the contour is defined on the complex cylinder in the $p$-adic case. We refer the reader to appendix \ref{BarnesSection} for more details.

\subsection{$p$-adic numbers and holographic correlators}
\label{p-adicRecap}

For a fixed prime number $p$, every non-zero $p$-adic number is given by a unique formal power series,
\eqn{}{
x = p^{v} \sum_{m=0}^\infty a_m p^m\,,
}
where the digits $a_m \in \{0,1,\ldots, p-1\}$ with $a_0 \neq 0$, and $v \in \mathbb{Z}$ is called  the $p$-adic valuation of $x$.  The $p$-adic norm, denoted $|\cdot|_p$, is then defined to be
\eqn{}{
|x|_p = p^{-v}\,,
}
with $|0|_p \equiv 0$.
The $p$-adic numbers, which form a field and are denoted $\mathbb{Q}_p$, are obtained as the completion of the rationals $\mathbb{Q}$ with respect to the $p$-adic norm $|\cdot|_p$, just like the field of real numbers is obtained as the completion of $\mathbb{Q}$ with respect to the absolute value norm. The $p$-adic norm obeys a stronger version of the triangle inequality; $|a+b|_p \leq \sup\{|a|_p, |b|_p\}$. This property is referred to as the ultrametricity of the $p$-adic norm.

In this paper, we will be working with the unique unramified field extension of $\mathbb{Q}_p$ of degree $n$, denoted $\mathbb{Q}_{p^n}$, which contains $\mathbb{Q}_p$ as a sub-field and may be viewed as an $n$-dimensional vector space over $\mathbb{Q}_p$. (Formally, setting $n=1$  recovers the base field $\mathbb{Q}_p$.) A unique ultrametric norm can be defined on the field extension, such that the field extension norm of any element $x \in \mathbb{Q}_p \subset \mathbb{Q}_{p^n}$   is precisely its $p$-adic norm $|x|_p$. Thus by abuse of notation, we will denote the norm in the field extension also by $|\cdot|_p$ and simply refer to it as the ``$p$-adic norm''.
For more details on the unramified field extension see, for instance, the review in section 2 of Ref.~\cite{Gubser:2016guj}.

\vspace{.5em} 
According to the $p$-adic AdS/CFT correspondence~\cite{Gubser:2016guj,Heydeman:2016ldy}, large $N$ conformal field theories living on a $p$-adic valued spacetime, for instance on the degree $n$ unramified extension of the $p$-adic numbers $\mathbb{Q}_{p^n}$, should admit a holographic description much like in the standard AdS$_{n+1}$/CFT$_n$ correspondence over the reals. 
Over the $p$-adics, the role of vacuum AdS space is played by the Bruhat--Tits tree $\mathcal{T}_{p^n}$  (also sometimes referred to as the Bethe lattice in the physics literature) for $p^n$ a positive integer power of a prime. $\mathcal{T}_{p^n}$ is a discrete $(p^n+1)$-regular graph without any cycles,  whose boundary at infinity is the projective line $\mathbb{P}^1(\mathbb{Q}_{p^n}) = \mathbb{Q}_{p^n} \cup \{\infty\}$. 
If we define the set of $p$-adic integers, $Z_{p^n} \equiv \{z\in \mathbb{Q}_{p^n} \,\,:\,\,|z|_p \leq 1\}$, then in the Poincar\'{e} patch picture~\cite{Gubser:2016guj}, each vertex on the Bruhat--Tits tree corresponds to a bulk point, and can be identified with a pair of coordinates $(z_0,z)$ where $z_0=p^\omega$ with $\omega \in \mathbb{Z}$ denoting the bulk depth (with more negative $\omega$ corresponding to vertices deeper in the bulk), and $z \in \mathbb{Q}_{p^n}$ denoting the boundary direction. Such an identification is highly non-unique, with any other pairing $(z_0,z^\prime)$ related to the original pairing $(z_0,z)$ via $z^\prime = z + z_0\mathbb{Z}_{p^n}$ also corresponding to the same bulk vertex on the Bruhat--Tits tree~\cite{Gubser:2016guj}.\footnote{This non-uniqueness in the description of the bulk coordinate in terms of the boundary coordinates encodes the relation between  bulk depth direction and boundary RG flow~\cite{Gubser:2016guj,Heydeman:2016ldy}.} In a more ``global picture'', any vertex on the Bruhat--Tits can be uniquely specified by choosing three points on the boundary $\mathbb{P}^1(\mathbb{Q}_p)$.

The simplest bulk action one can write down on the Bruhat--Tits tree is the free lattice action for a real-valued bulk scalar field $\phi$ (defined on the vertices of the tree) of mass-squared $m_\Delta^2$ (and conformal dimension $\Delta$) which lives on the vertices of the Bruhat--Tits tree,
\eqn{FreeAction}{
S_{\text{kin}}=\sum_{\left<(z_0,z)(w_0,w)\right>}{1 \over 2}(\phi_{(z_0,z)}-\phi_{(w_0,w)})^2+\sum_{(z_0,z) \in \mathcal{T}_{p^n}}
\frac{1}{2} m^2_\Delta\phi_{(z_0,z)}^2\,,
}
where the first sum is taken over all pairs of neighbouring vertices on the tree (i.e.\ over all edges), while the second sum is over all vertices of the tree. 
Further, the classic mass-dimension relation takes the following form in $p$-adic AdS/CFT~\cite{Gubser:2016guj}
\eqn{massDimension}{
m_{\Delta}^2 = {-1 \over \zeta_p(-\Delta)\zeta_p(\Delta-n)}\,.
}
To get a theory with non-trivial correlators, it is necessary to introduce interactions. In a perturbative expansion in the coupling constant, the leading order contribution to the correlators can be depicted graphically as tree-diagrams (not to be confused with the underlying space which is itself a tree), one important class of which is contact diagrams. Letting the external operators in a contact diagram carry different scaling dimensions presents little extra difficulty, so we will consider a theory with $\mathcal{N}$ different bulk scalar fields $\phi^i$ of mass $m_{\Delta_i}$ and conformal dimension $\Delta_i$ obeying \eno{massDimension}, and contact interaction terms of the type
\eqn{InteractionTerm}{
\sum_{(z_0,z)\in\mathcal{T}_{p^n}}\prod_{i=1}^\mathcal{N}\phi^i_{(z_0,z)}\,,
}
for $\mathcal{N} \geq 3$. 
This interaction \eno{InteractionTerm} represents the $p$-adic analog of a local $\mathcal{N}$-point interaction term in continuum AdS space of the form $\big(\phi_{\Delta_1}(x) \ldots \phi_{\Delta_{\mathcal{N}}}(x)\big)$, where $\phi_\Delta$ is a bulk field of conformal dimension $\Delta$. We omit overall coupling constant factors.

\begin{figure}
\centering{
\includegraphics[height=18ex]{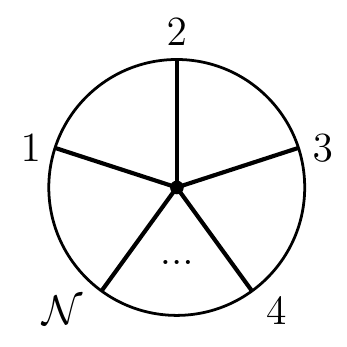} \hspace{4em}
\includegraphics[height=15ex]{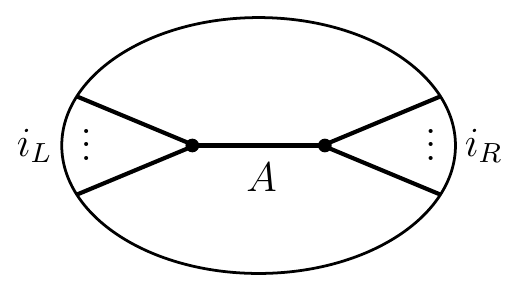}
}
\caption{Left: $\mathcal{N}$-point bulk contact diagram. Right: Arbitrary-point bulk exchange diagram.}
\label{fig:ContactSingle}
\end{figure}

$\mathcal{N}$-point bulk contact diagrams (see figure \ref{fig:ContactSingle}) are given by the product of $\mathcal{N}$ bulk-to-boundary propagators from $\mathcal{N}$ distinct boundary points $x_i$ to the same bulk point of integration $(z_0,z)$, as follows
\eqn{contact}{
\mathcal{A}^{\text{contact}}(x_i)
=
\sum_{(z_0,z)\in \mathcal{T}_{p^n}} 
\prod_{i=1}^\mathcal{N} {K}_{\Delta_i}(z_0,z;x_i)\,,
}
where ${K}_{\Delta_i}$ are the bulk-to-boundary propagators discussed in section \ref{GKETC}.
The bulk point $(z_0,z)$ in \eno{contact} is integrated over the entire bulk space. On the Bruhat--Tits tree, such integrations reduce to discrete summations over the vertices of the tree; see the discussion around \eno{SumToInt} in section \ref{PADICSETUP} for the connection between a continuum integral prescription and the tree-summation.

Another class of bulk diagrams are the exchange diagrams, those which admit exactly one single-trace exchange of dimension $\Delta_A$ (see figure \ref{fig:ContactSingle}), given by
\eqn{OneExchange}{
\mathcal{A}^{\text{exch}}=\!\!\!\!
\sum_{(z^L_0,z^L) \in \mathcal{T}_{p^n}} \sum_{(z^R_0,z^R)\in \mathcal{T}_{p^n}}\!\!\!\bigg(\prod_{i_L}{K}_{\Delta_{i_L}}(z^L_0,z^L;x_{i_L})\bigg) G_{\Delta_A}(z^L,z^L_0;z^R,z^R_0) 
\bigg(\prod_{i_R}K_{\Delta_{i_R}}(z^R_0,z^R;x_{i_R})\bigg),
}
where the product over the index $i_L$ ($i_R$) runs over all external legs to the left (right) of the single-trace exchange depicted in figure \ref{fig:ContactSingle}. Here $G_\Delta$ is the bulk-to-bulk propagator for a scalar field of conformal dimension $\Delta$, and is discussed later in section \ref{GKETC}.
Such $p$-adic position space amplitudes were first computed in the case of the three- and four-point contact diagrams and the four-point exchange diagram in Refs.~\cite{Gubser:2016guj,Gubser:2017tsi} and represent the current state-of-the-art in $p$-adic AdS/CFT. 

 For higher point bulk Feynman diagrams, such as the five-point contact diagram and exchange diagrams with one or two internal lines,  geodesic bulk diagram techniques of Ref.~\cite{Hijano:2015zsa} adapted to the $p$-adics~\cite{Gubser:2017tsi}, together with various propagator identities of Ref.~\cite{Gubser:2017tsi} can be used to obtain closed-form position space expressions, though such expressions become tedious to write down when going beyond five points. 
However, it is well known that in standard AdS/CFT, Mellin space amplitudes assume much simpler forms. Moreover, the complexity of the expressions does not generically increase with the number of external insertion points. In this paper we find that the same observation holds true over the $p$-adics. Thus $p$-adic Mellin amplitudes, as introduced in section \ref{MELLINSPACE}, provide a convenient framework for studying arbitrarily complicated bulk diagrams.

\subsection{Summary and organization}
\label{SUMMARY}

 The new results of this paper comprise the formulation and the first principles computation of $p$-adic Mellin amplitudes.
  We have already proposed the definition of $p$-adic Mellin amplitudes in section \ref{MELLINSPACE}. Before moving to the actual computation of such amplitudes, we show using a simple example in section \ref{EXAMPLE4PT}, how the $p$-adic Mellin formula \eqref{pMel} works --- exactly which contours the Mellin variables are integrated over, and how the position space amplitude is recovered given the Mellin amplitude.
 
 To obtain $p$-adic Mellin amplitudes, which is our main goal, we start with position space amplitudes such as those written in \eno{contact} and \eno{OneExchange} and use various manipulations to rewrite them in the form given in \eno{pMel}, from which we can simply read off the Mellin amplitudes. 
Two key ingredients in this procedure will be: (a) the $p$-adic version of the well-known Schwinger parameter trick, which allows one to carry out bulk summations, and (b) the $p$-adic analog of the Gaussian function, the so-called characteristic function. Both these ingredients are the subject of  section \ref{PADICSETUP}.
The computation of the $p$-adic scalar ${\cal N}$-point  contact Mellin amplitude $\mathcal{M}^{\text{contact}}$ is similar in spirit to the analogous calculation over the reals and is detailed in section \ref{CONTACT}. The end result of a non-trivial calculation is that
\eqn{padicContact}
{
\mathcal{M}^{\text{contact}}=\zeta_p\left(\sum\Delta_i-n\right),
}
where $\sum \Delta_i$ represents the sum over all external dimensions.
As in the real case, the contact amplitude is a constant, i.e.\ independent of Mellin variables $\gamma_{ij}$. 
We note further that for a suitable normalization of the bulk-to-boundary propagators,\footnote{\label{fn:realNormalization}We have chosen the normalizations for the bulk-to-bulk and bulk-to-boundary propagators in line with the choice we make later for the corresponding $p$-adic propagators  in \eno{NormalizationChoice} but different from the convention used in Ref.~\cite{Penedones:2010ue}. Specifically, we have here
\eqn{Greal}{
G_\Delta(Z,W)= {\zeta_\infty(2\Delta) \over (Z-W)^{2\Delta}} {}_2F_1\left(\Delta, \Delta-{n\over 2}+{1\over 2};2\Delta-n+1;-{4\over (Z-W)^2} \right) \,,
}
where $Z, W \in \mathbb{M}^{n+1,1}$ are embedding space coordinates in $(n+2)$-dimensional Minkowski space satisfying $Z^2=W^2=-1$, and 
\eqn{Kreal}{
K_\Delta(Z, P) = {\zeta_\infty(2\Delta) \over (-2P \cdot Z)^\Delta}\,,
}
where $P \in \mathbb{M}^{n+1,1}$ and $P^2=0$, so that $P$ can be thought of as a coordinate on the conformal boundary of the AdS hyperboloid. 
}
and for the definition of ${\cal M}$ as given in \eno{ArcMellin}  (except with the factors of $\Gamma(\gamma_{ij})$  replaced by the corresponding factors of $\zeta_\infty(2\gamma_{ij})$ in the definition \eno{ArcMellin}), the real contact Mellin amplitude  is given by \cite{Penedones:2010ue}
\eqn{realContact}
{
\mathcal{M}^{\text{contact}} = \frac{1}{2}\,\zeta_\infty\left(\sum\Delta_i-n\right),
}
where the local zeta function $\zeta_\infty$ was defined in \eno{zetainfty}.
Equations \eno{padicContact} and \eno{realContact} provide yet another example of how, for reasons not yet fully understood, many formulas in $p$-adic AdS/CFT look almost exactly identical to their real counterparts, when expressed in terms of the right functions.\footnote{\label{fn:FactorTwo} The factor of $2$ mismatch between the real and $p$-adic results also manifests itself in position space expressions~\cite{Gubser:2016guj}, and may be thought of as resulting from the choice of normalization of integration measures: The $p$-adic Haar measure is conventionally normalized such that $\int_{|x|_p\leq 1}dx=1$, while over the reals $\int_{|x|\leq 1}dx=2$. }

\begin{figure}
\centering
\eqn{}
{  \begin{matrix}
\includegraphics[height=22ex]{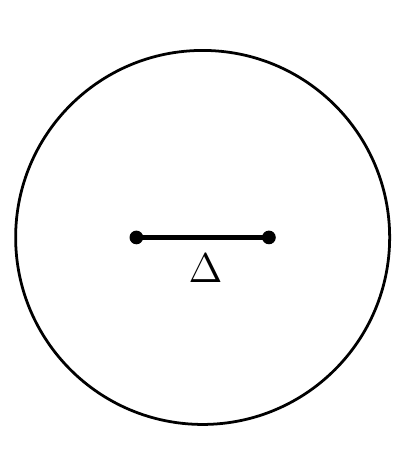}
 \end{matrix}=\int {dc \over 2\pi i}\,\,f_\Delta(c)\int_{\partial\mathcal{T}_{p^n}}dx\,
 \begin{matrix}
 \includegraphics[height=22ex]{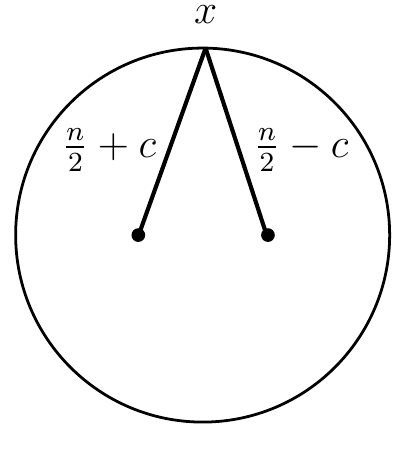}
  \end{matrix}
  \nonumber
  }
\caption{The split representation \eno{HarmExp}.}
\label{fig:SplitRep}
\end{figure}

For bulk diagrams with one or more internal lines, in the standard AdS/CFT setup it is useful to apply the split representation~\cite{Penedones:2010ue} (also referred to as the spectral representation or the harmonic expansion) of the bulk-to-bulk-propagator. 
The split representation re-expresses the bulk-to-bulk propagator as a contour integral over a product of two bulk-to-boundary propagators connected to the same boundary point, which  is to be integrated over the whole boundary, thereby permitting one to recast any tree-level (or even higher-loop) diagram with internal exchanges as a multi-dimensional contour integral over a product of appropriate contact interactions.
In section \ref{HARMEXPAND} we derive the following $p$-adic version of the split representation (see also figure \ref{fig:SplitRep}),
\eqn{HarmExp}
{
G_{\Delta}(z_0,z;w_0,w) &= {\nu_p \over 2} \, \zeta_p(2\Delta-n)
 \int_{-\frac{i\pi}{\log p}}^{\frac{i\pi}{\log p}}\, {dc \over 2\pi i/(2\log p)}\,\frac{1}{\zeta_p(2c)\zeta_p(-2c)} {1 \over m^2_{\Delta}-m^2_{n/2-c}} \cr 
& \quad 
 \times \int_{\partial {\cal T}_{p^n}}dx\,K_{\frac{n}{2}-c}(z_0,z;x)K_{\frac{n}{2}+c}(w_0,w;x)\,, 
}
where
\eqn{nupDef}{
2\nu_p \equiv p^{\Delta_+} - p^{\Delta_-} = { p^\Delta \over \zeta_p(2\Delta-n)} \quad\qquad (\Delta_+ = \Delta\,, \Delta_- = n-\Delta) \,,
}
where the bulk-to-bulk and bulk-to-boundary propagators $G_\Delta$ and $K_\Delta$ are given later in \eno{bulk-to-bulk} and \eno{bulk-to-boundary}, the conformal boundary is given by $\partial {\cal T}_{p^n} = \mathbb{P}^1(\mathbb{Q}_{p^n})$, and $m^2_\Delta$ obeys \eno{massDimension}.

Interestingly, by comparison, the real analog of \eno{HarmExp} (in embedding space) is given by~\cite{Penedones:2010ue}
\eqn{HarmExpReal}{
G_{\Delta}(Z,W) &= {\nu_\infty\over 2}\, \zeta_\infty(2\Delta-n)
 \int_{-i\infty}^{i\infty}\, {dc \over 2\pi i}\,\frac{1}{\zeta_\infty(2c)\zeta_\infty(-2c)} {1 \over m^2_{\Delta}-m^2_{n/2-c}} \cr 
& \quad 
 \times \int_{\partial {\rm AdS}}dP\,K_{\frac{n}{2}-c}(Z,P)K_{\frac{n}{2}+c}(W,P)\,, 
}
where 
\eqn{nuinftyDef}{
2\nu_\infty \equiv \Delta_+-\Delta_- = 2\Delta-n  \quad \qquad (\Delta_+ = \Delta \,, \Delta_- = n-\Delta) \,,
}
and now $m_\Delta^2 = \Delta(\Delta-n)$.\footnote{For a discussion on the relation between the overall factors $\nu_p$ and $\nu_\infty$ see sections\ 5.1-5.2 of Ref.~\cite{Gubser:2016guj}, where precisely the same factors make an appearance.} 
We have chosen to express \eno{HarmExpReal} in a non-standard way, using the local zeta function $\zeta_\infty$ and the mass-squared of the bulk scalar field to emphasize the similarity with the corresponding $p$-adic result \eno{HarmExp}. 
However it is worth noting that \eno{HarmExpReal} is simply a repackaging of e.g.\ equation\ (121) of Ref.~\cite{Penedones:2010ue} with the choice of normalization given in footnote \ref{fn:realNormalization}.

With \eno{HarmExp} in hand, we proceed in sections \ref{singleSection}-\ref{tripleSection} to calculate the $p$-adic Mellin amplitudes for arbitrary-point tree-level diagrams with one, two or three internal lines. For example, we show in section \ref{singleSection} that the Mellin amplitude for the diagram \eno{OneExchange} is given in the so-called Mellin-Barnes contour integral representation by
\eqn{padicExchExp}{
{\cal M}^{\rm exch} = 2\nu_p\, { \zeta_p(2\Delta-n) \over \zeta_p(\sum_{i_L}\Delta_{i_L} - s) \zeta_p(\sum_{i_R}\Delta_{i_R} - s)} \int_{-{i \pi \over \log p}}^{{i \pi \over \log p}} {dc \over 2\pi i /(2\log p)} {\ell_{n\over 2}(c) \ell_{n\over 2}(-c) \over m^2_\Delta - m^2_{n/2-c}}\,,
}
where $\nu_p$ is given in \eno{nupDef}, $m^2_\Delta$ is given by \eno{massDimension}, and we have defined
\eqn{ellDef}{
\ell_{n\over 2}(c) \equiv {\zeta_p(c+n/2-s)\zeta_p(\sum_{i_L}\Delta_{i_L}+c-n/2)\zeta_p(\sum_{i_R}\Delta_{i_R}+c-n/2) \over 2\,\zeta_p(2c)} \,.
}
The Mandelstam-like variable $s$ is defined to be
\eqn{sExch}{
s \equiv \sum_{i_L}\Delta_{i_L} - 2 \sum_{\substack{j < k\\ j,k \in i_L}} \gamma_{jk} = \sum_{i_R}\Delta_{i_R} - 2 \sum_{\substack{j<k\\ j,k \in i_R}} \gamma_{jk} \,.
}
The Mellin amplitude for the real analog of \eno{OneExchange} takes an almost identical form in its Mellin Barnes representation~\cite{Penedones:2010ue}, and can be written as
\eqn{realExchExp}{
{\cal M}^{\rm exch} = 
\nu_\infty\,    { \zeta_\infty(2\Delta-n) \over \zeta_\infty(\sum_{i_L}\Delta_{i_L} - s) \zeta_\infty(\sum_{i_R}\Delta_{i_R} - s)} \int_{-i\infty}^{i \infty} 
{dc \over 2\pi i} {\ell_{n\over 2}(c) \ell_{n\over 2}(-c) \over m^2_\Delta - m^2_{n/2-c}}\,,
}
where $m^2_\Delta = \Delta(\Delta-n)$, $\nu_\infty$ is given in \eno{nuinftyDef}, $\ell_{n\over 2}(c)$ is defined exactly as in \eno{ellDef} except with $\zeta_p$ replaced by $\zeta_\infty$, 
and $s$ is given by \eno{sExch}. The amplitude \eno{realExchExp} is  simply a rewriting of equation (46) in Ref.~\cite{Penedones:2010ue} (suppressing overall coupling constant factors) in terms of $\zeta_\infty$ and $m^2_\Delta$, except with a choice of normalization for propagators as noted in footnote \ref{fn:realNormalization} and a choice of normalization for ${\cal M}$ as prescribed by a modification of \eno{ArcMellin} where the explicit factors of $\Gamma(\gamma_{ij})$ have been replaced by the corresponding factors of $\zeta_\infty(2\gamma_{ij})$. 
The pole structure of the $p$-adic and real Mellin amplitudes in the Mandelstam variable $s$, in \eno{padicExchExp} and \eno{realExchExp} respectively,  takes a particularly similar form; the only difference arises from the fact that the $\zeta_\infty$ functions have a semi-infinite sequence of poles while  only the first pole in this semi-infinite sequence survives as a pole of $\zeta_p$. 
This observation captures the essence of the general wisdom that $p$-adic and real amplitudes are closely related, yet the $p$-adic case is decidedly simpler. Indeed, equations \eno{padicContact}-\eno{realExchExp} already provide strong evidence for a connection between real and $p$-adic Mellin amplitudes.

 We close the paper with final comments and future directions in section \ref{DISCUSSION}.

\section{From Mellin Space to Position Space}
\label{EXAMPLE4PT}

As noted earlier, for large $N$ CFTs, the usual (real) Mellin amplitude for a bulk contact diagram of scalar primaries is simply a constant, owing to the fact that there are no single-trace operator exchanges in the intermediate channel, while the double-trace contribution is precisely reproduced from the poles of the Euler gamma function factors in \eno{ArcMellin}. 
For the same reason, it is reasonable to expect that the $p$-adic Mellin amplitude for the same contact diagram be simply a constant, with the poles of the local zeta function factors in \eno{pMel} reproducing the double-trace contribution.

\begin{figure}
\centering
\begin{subfigure}[b]{0.35\textwidth}
        \includegraphics[width=\textwidth]{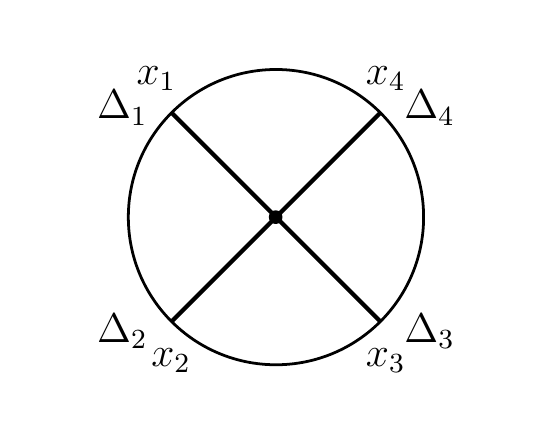}
        \caption{}
         \label{fig:4ptContacta}
\end{subfigure}
\hspace{20mm}
\begin{subfigure}[b]{0.35\textwidth}
        \includegraphics[width=\textwidth]{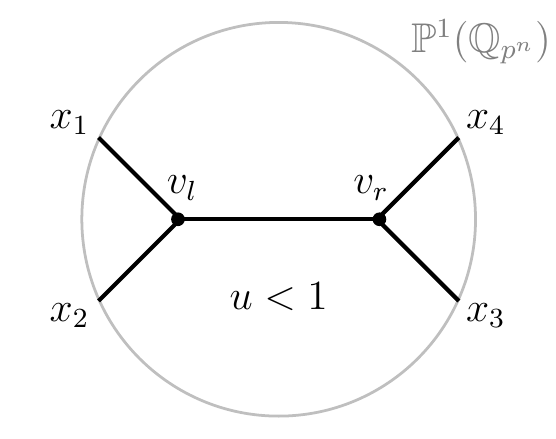}
        \caption{}
         \label{fig:4ptContactb}
\end{subfigure}

\caption{(a) The bulk 4-point contact Feynman diagram for scalar fields with scaling dimensions $\Delta_i$. (b) The coordinate configuration on the Bruhat--Tits tree. Solid lines are geodesics on the Bruhat–Tits tree, tracing the path joining together the four points on the boundary of the tree, which is the projective line over the degree $n$ unramified extension of $\mathbb{Q}_p$. The figure is drawn for $u < 1$ where $u, v$ are defined in \eno{uvDef}. For the $u=v=1$ configuration, the vertices on the Bruhat--Tits tree, labeled $v_l$ and $v_r$, become coincident.}
\label{fig:4ptContact}
\end{figure}
For definiteness, let us specialize to the case of the four-point contact diagram with (external) scaling dimensions $\Delta_1$, $\Delta_2$, $\Delta_3$, and $\Delta_4$ (see figure \ref{fig:4ptContacta}). 
The position space expression for this diagram was first computed in Ref.~\cite{Gubser:2017tsi} in the context of $p$-adic AdS/CFT. 
The four-point contact diagram on the Bruhat--Tits tree is given by
\eqn{4ptPosition}{
{\cal A}(\{x_i\}) = \sum_{a\in \mathcal{T}_{p^n}}\prod_{i=1}^4\hat{K}_{\Delta_i}(a;x_i)\,,
}
where, just for this section, we use the unnormalized bulk-to-boundary propagators $\hat{K}_{\Delta_i}$ which are discussed in section \ref{GKETC}, and label the bulk point $a=(z_0,z) \in T_{p^n}$ for appropriately chosen $(z_0,z)$. 
In this section, we will reproduce the position space result for the four-point contact diagram~\cite{Gubser:2017tsi} starting from \eno{pMel} and the assumption that the $p$-adic Mellin amplitude for the contact diagram is a Mellin variable independent constant, ${\cal M} (\gamma_{ij}) = {\cal M}$.

We begin by choosing $\gamma_{12}$ and $\gamma_{14}$ to be the ${4\times (4-1) \over 3}= 2$ independent Mellin variables, so that the remaining Mellin variables are given by
\eqn{MellinVars4pt}
{
&\gamma_{13}=\Delta_1-\gamma_{12}-\gamma_{14} \cr
&\gamma_{23}=\frac{\Delta_{23,14}}{2}+\gamma_{14}\cr
&\gamma_{24}=\frac{\Delta_{124,3}}{2}-\gamma_{12}-\gamma_{14}\cr
&\gamma_{34}=\frac{\Delta_{34,12}}{2}+\gamma_{12}\,,
}
where we have adopted the short-hand
\eqn{DeltaijkComma}
{
\Delta_{i_1...i_k,i_{k+1}...i_{l}}\equiv\sum_{j=1}^k\Delta_{i_j}-\sum_{j=k+1}^l\Delta_{i_j}\,.
}
The expressions in \eno{MellinVars4pt} are obtained by solving the constraints \eno{MellinVarConstraints}. Further, we write
\eqn{xijDef}{
x_{ij} \equiv x_i - x_j\,.
} 
The Mellin representation \eno{pMel}
\eqn{pMel4pt}
{
\mathcal{A}(\{x_i\})= {\cal M} \int [d\gamma] \prod_{1\leq i<j\leq 4}\zeta_p(2\gamma_{ij})|x_{ij}|_p^{-2\gamma_{ij}}\,,
}
then takes the explicit form
\eqn{4ptexpanded}
{
{\cal A} &= {\cal M} |x_{13}|_p^{-2\Delta_1}|x_{23}|_p^{-\Delta_{23,14}}|x_{24}|_p^{-\Delta_{124,3}}|x_{34}|_p^{-\Delta_{34,12}}  \int \frac{d\gamma_{14}}{{\pi i\over \log p}} \, 
\zeta_p\left(2\gamma_{14}\right) 
\zeta_p\left(2\gamma_{14}+\Delta_{23,14}\right) v^{-2\gamma_{14}} \cr
& \times\int\frac{d\gamma_{12}}{{\pi i\over \log p}} \,
\zeta_p\left(2\gamma_{12}\right)
\zeta_p\left(2\gamma_{12}+\Delta_{34,12}\right)
\zeta_p\left(2\Delta_1-2\gamma_{12}-2\gamma_{14}\right)
\zeta_p\left(\Delta_{124,3}-2\gamma_{12}-2\gamma_{14}\right) u^{-2\gamma_{12}} \,.
}
Here we have defined the conformally invariant cross ratios
\eqn{uvDef}{
u \equiv \left|\frac{x_{12}x_{34}}{x_{13}x_{24}}\right|_p \qquad
v \equiv \left|\frac{x_{14}x_{23}}{x_{13}x_{24}}\right|_p\,.
}
Because of the ultrametricity of the $p$-adic norm, we can assume without loss of generality that the indices of the external legs are labeled such that $u\leq 1$ and $v=1$ (see figure \ref{fig:4ptContactb}).\footnote{If $u\geq 1$ we can interchange indices 2 and 3 to make $u\leq 1$. Let $a=\frac{x_{12}x_{34}}{x_{13}x_{24}}$ and $b=\frac{x_{14}x_{23}}{x_{13}x_{24}}$ such that $u=|a|_p$ and $v=|b|_p$. It is straightforward to check that $a+b=1$. But for any triplet of $p$-adic numbers $\{ a, b, a+b\}$, it holds true that the $p$-adic norms of two of them must be equal and cannot be smaller than the norm of the third. Since we've enforced $|a|_p\leq 1$, we must have either that $|b|_p=1$ or that $|a|_p=1$ and $|b|_p\leq 1$. In the latter case we can interchange indices $2$ and $4$ to make $|a|_p\leq 1$ and $|b|_p=1$.}

\begin{figure}
\centering{
\includegraphics[height=22ex]{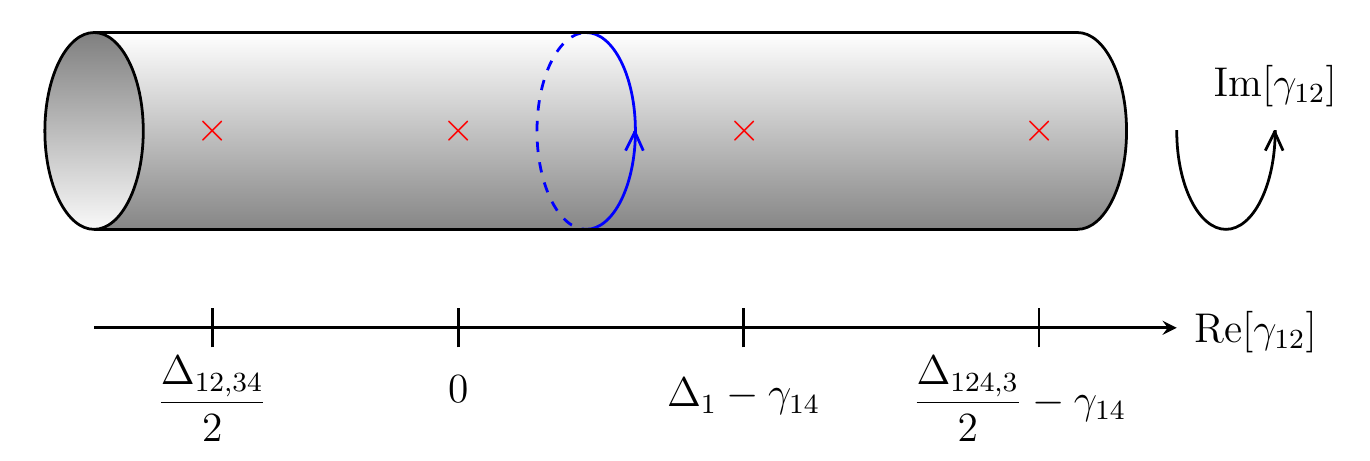}
}
\caption{Integration contour for $\gamma_{12}$ for computing the position space 4-point contact amplitude starting from its Mellin representation \eno{pMel4pt}. The circumference of the cylinder is $\frac{\pi}{\log p}$.}
\label{fig:4ptContour}
\end{figure}
To evaluate \eqref{4ptexpanded}, we first need to describe the contour prescription for the inside integral over $\gamma_{12}$. 
The appropriate integration contour is depicted in figure \ref{fig:4ptContour}: it is a circular contour along the periodic imaginary direction wrapping around the cylinder, with the poles at $\Delta_{12,34}/2$ and 0 on one side (on the left in figure \ref{fig:4ptContour}) and the poles at $\Delta_1-\gamma_{14}$ and $\Delta_{124,3}-\gamma_{14}$ on the other  (on the right in figure \ref{fig:4ptContour}).\footnote{For definiteness, the figure has been drawn for the case where $\Delta_{12,34}<0$ and $\Delta_1<\Delta_{124,3}/2$, but we do not assume that in the calculation. However, we do require $0\,, \Delta_{12,34}/2 < \Re[\Delta_1-\gamma_{14}], \Re[\Delta_{124,3}/2-\gamma_{14}]$.} 
More precisely, thinking of the cylinder as $\mathbb{R} \times S^1$, the $S^1$-direction is identified with the imaginary part of $\gamma_{12}$, with the $\mathbb{R}$-direction identified with the real part of $\gamma_{12}$.  
The poles are obtained by setting the arguments of the local zeta functions in \eno{4ptexpanded} to zero. The dichotomy in the position of the poles originates from looking at the arguments of the local zeta function $\zeta_p$ in the second line of \eno{4ptexpanded}: All poles originating from a local zeta function whose argument contains $\gamma_{12}$ with a negative sign lie on one side of the $\gamma_{12}$ integration contour, while poles coming from local zeta functions which contain $\gamma_{12}$ with a positive sign lie on the other side. Note that a consequence of this prescription is that if one translates the integral \eqref{4ptexpanded} into its real analog by letting the radius of the cylindrical manifold tend to infinity and replacing the $p$-adic local zeta function $\zeta_p(z)$ with the local zeta function at infinity, $\zeta_\infty(z)$, then the integration contour will lie entirely to the left or right of the semi-infinite sequences of poles arising from the Euler gamma functions.

As long as the circular contour encounters no poles, we can freely slide it along the cylinder without affecting the integral. But in moving the contour past poles, we pick up contributions from the residues of the poles. Specifically, we shift the contour to Re$[2\gamma_{12}]=-\infty$ at the cost of $2\pi i$ times the sum of the residues at $\Delta_{12,34}/2$ and 0. Since $u\leq 1$, the boundary integral vanishes and carrying out the $\gamma_{12}$ integral of \eqref{4ptexpanded} leaves us with
\eqn{expanded2}
{
{\cal A} &= {\cal M} |x_{13}|_p^{-2\Delta_1}|x_{23}|_p^{-\Delta_{23,14}}|x_{24}|_p^{-\Delta_{124,3}}|x_{34}|_p^{-\Delta_{34,12}}\cr
& \times \int_{-i\infty+|\epsilon|}^{i\infty+|\epsilon|}\frac{d\gamma_{14}}{{\pi i \over \log p}}
\zeta_p\left(2\gamma_{14}\right)
\zeta_p\left(2\gamma_{14}+\Delta_{23,14}\right) \bigg[
\zeta_p\left(\Delta_{34,12}\right)\zeta_p(2\Delta_1-2\gamma_{14})\zeta_p\left(\Delta_{124,3}-2\gamma_{14}\right) \cr 
& \qquad +\zeta_p\left(-\Delta_{34,12}\right)\zeta_p\left(\Delta_{134,2}-2\gamma_{14}\right)\zeta_p\left(2\Delta_4-2\gamma_{14}\right)u^{\Delta_{34,12}}\bigg]\,,
}
where $\epsilon$ is any small number such that the integration contour around the cylindrical manifold has the poles at $0$ and $\frac{\Delta_{14,23}}{2}$ on one side and the poles at $\Delta_1$, $\Delta_4$, $\frac{\Delta_{124,3}}{2}$, and $\frac{\Delta_{134,2}}{2}$ on the other.
We next carry out the $\gamma_{14}$ integral, e.g. by summing over the residues at $0$ and $\frac{\Delta_{14,23}}{2}$, leaving us with
\eqn{}
{
{\cal A} &= {\cal M} \left|\frac{x_{24}}{x_{14}}\right|_p^{\Delta_{1,2}}\left|\frac{x_{14}}{x_{13}}\right|_p^{\Delta_{3,4}}\frac{1}{|x_{34}|_p^{\Delta_{34,}}|x_{12}|_p^{\Delta_{12,}}}
\bigg[\frac{\zeta_p(2\Delta_1)\zeta_p(2\Delta_2)\zeta_p(\Delta_{34,12})\zeta_p(\Delta_{123,4})\zeta_p(\Delta_{124,3})}{\zeta_p(2\Delta_{12,})}u^{\Delta_{12,}} \cr 
& \qquad +(1\leftrightarrow 4,2\leftrightarrow 3)\bigg]\,,
}
where we remind the reader that, for instance, $\Delta_{12,} = \Delta_1 + \Delta_2$ and $\Delta_{123,4} = \Delta_1+\Delta_2+\Delta_3-\Delta_4$, while $x_{12} = x_1-x_2$.
This expression reproduces the precise position space dependence of the four-point contact amplitude computed via geodesic bulk diagram techniques (a.k.a.\ geodesic Witten diagram techniques)~\cite{Gubser:2017tsi}, and in fact matches the overall normalization as well if we choose
\eqn{Mfactor}
{
\mathcal{M}=\frac{\zeta_p(\Delta_{1234,}-n)}{\zeta_p(2\Delta_1)\zeta_p(2\Delta_2)\zeta_p(2\Delta_3)\zeta_p(2\Delta_4)}\,.
}
We note that this result differs from \eno{padicContact} in its overall normalization due to the fact that we used the unnormalized bulk-to-boundary propagators in \eno{4ptPosition}.
Thus to summarize, we have shown that
\eqn{4ptResult}
{
{\cal A}(\{x_i\}) = \sum_{a\in \mathcal{T}_{p^n}}\prod_{i=1}^4\hat{K}_{\Delta_i}(a;x_i)=
\frac{\zeta_p(\sum_{i=1}^4\Delta_i-n)}{\prod_{i=1}^n\zeta_p(2\Delta_i)}\int [d\gamma] \prod_{1\leq i<j\leq 4}\zeta_p(2\gamma_{ij})|x_{ij}|_p^{-2\gamma_{ij}}\,.
}
While in this section we reproduced the position space amplitude simply by guessing the $p$-adic Mellin amplitude by analogy with the real Mellin amplitude, we will   derive from first principles the generalization of \eno{4ptResult} to arbitrary-point contact diagrams in section \ref{CONTACT}.

\section{Preliminaries: The $p$-adic Toolbox}
\label{PADICSETUP}

Many of the steps involved in computing $p$-adic Mellin amplitudes  closely mirror corresponding steps in computing real Mellin amplitudes, but there also occur several subtleties that are peculiar to working with the $p$-adic numbers. In this section we set up some notation we will be adopting in the following and present and explain various $p$-adic computational tools and techniques that will prove useful in deriving explicit expressions for $p$-adic Mellin amplitudes. We end the section with a presentation of the bulk-to-bulk and bulk-to-boundary propagators in $p$-adic AdS/CFT.

\subsection{The characteristic function}

Over $p$-adics, the role of the Gaussian function is played by  the characteristic function of  $p$-adic integers ${\mathbb{Z}}_{p^n}$, which were defined in section \ref{p-adicRecap}. The characteristic function is denoted $\gamma_p$ and is defined as follows,
\eqn{gammafun}
{\gamma_p(x) \equiv \begin{cases}
1 \quad \text{ for } x \in \mathbb{Z}_{p^n}, \\
0 \quad \text{ otherwise. }
\end{cases}}
In other words $\gamma_p(x) = 1$ iff $|x|_p \leq 1$; otherwise it vanishes. This function features prominently in the rest of the paper, so we briefly discuss some of its properties here.

As demonstrated e.g.\ in Ref.~\cite{Gubser:2016guj}, the characteristic function, just like the Gaussian over the reals is its own Fourier transform. However, it factorizes significantly differently than the Gaussian, namely as  
\eqn{GammaFactorization}
{
\gamma_p(x_1)...\gamma_p(x_\mathcal{N}) = \gamma_p\big((x_1,...,x_{\cal N})_s\big),
}
where 
\eqn{supVariable}{
(x_1,\ldots,x_\mathcal{N})_s \equiv \begin{cases} x_1 & {\rm if\ } |x_1,\ldots,x_\mathcal{N}|_s = |x_1|_p \\ 
										x_2 & {\rm if\ } |x_1,\ldots,x_\mathcal{N}|_s = |x_2|_p \\
                                        \ldots \\
                                        x_{\cal N} & {\rm if\ } |x_1,\ldots,x_\mathcal{N}|_s = |x_{\cal N}|_p \\
                            \end{cases}\,,
}
with the added stipulation that when multiple cases above are simultaneously true, $(x_1,
\ldots,x_\mathcal{N})_s$ can be set equal to any element from the set $\{ x_j : |x_1,\ldots,x_{\cal N}|_s = |x_j|_p, 1\leq j \leq \mathcal{N}\}$. Thus $(x_1,\ldots,x_\mathcal{N})_s$ is ill-defined as a function from $\left(\mathbb{Q}_{p^n}\right)^{\mathcal{N}} \to \mathbb{Q}_{p^n}$. However, in this paper such $(x_1,\ldots,x_\mathcal{N})_s$ will only appear in the argument of the characteristic function, and $\gamma_p((x_1,\ldots,x_\mathcal{N})_s)$ is well-defined since it only depends on the norm of its argument. The property \eno{GammaFactorization} can be  verified directly from the definition \eno{gammafun}.

Another useful property of $\gamma_p$, which follows from the ultrametricity of the $p$-adic norm, is that for any $p$-adic number $x\in \mathbb{Q}_{p^n}$, and any $p$-adic integer $z\in \mathbb{Z}_{p^n}$,\footnote{Due to the $\mathbb{Z}_{p^n}$ invariance of the characteristic function as exhibited in \eno{GammaAddition}, $\gamma_p$ really is a function on $\mathbb{Q}_{p^n}/\mathbb{Z}_{p^n}$.}
\eqn{GammaAddition}{
\gamma_p(x+z)=\gamma_p(x)\,,
}
that is, it is invariant under translations by $p$-adic integers.

\begin{figure}
\centering{
\includegraphics[height=22ex]{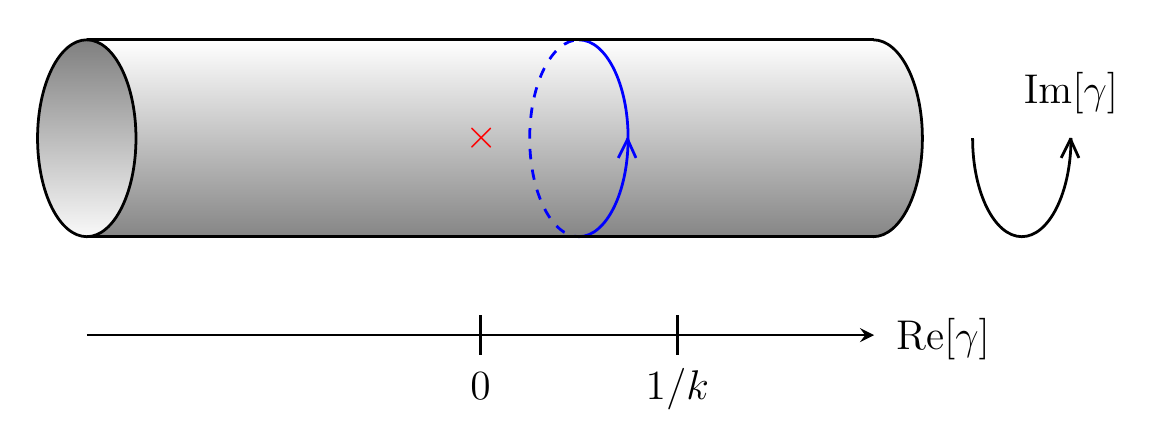}
}
\caption{The characteristic function $\gamma_p(x)$ can be expressed in terms of a closed contour integral running around a cylinder with a circumference of $\frac{2\pi}{k\log p}$.}
\label{fig:gammaContour}
\end{figure}

The characteristic function admits a representation in terms of a contour integral as follows, 
\eqn{gammaRep}
{\gamma_p(x)=\frac{k\log p}{2\pi i}\int_{\epsilon-\frac{i\pi}{k\log p}}^{\epsilon+\frac{i\pi}{k\log p}}d\gamma\,\frac{\zeta_p(k\gamma)}{|x|_p^{k\gamma}} \qquad k>0\,,}
where $k$ is a positive number and $\epsilon$ is a real number between $0$ and $1/k$. 
Because the integrand is periodic in the imaginary direction with periodicity ${2\pi}/(k\log p)$, the contour can be thought of as a closed loop around a cylinder as shown in figure \ref{fig:gammaContour}. 
On the cylinder, $\zeta_p(k\gamma)$ has but one pole, namely the simple pole at $\gamma=0$. To prove \eno{gammaRep}, we observe first that when $|x|_p \leq 1$, the integrand dies off if $\Re [\gamma]\rightarrow -\infty$. So we shift the contour left to $\Re [\gamma]\rightarrow -\infty$ where it vanishes, but we pick up  the residue at $\gamma=0$, which combines with the pre-factor to yield unity. 
For $|x|_p > 1$ the integrand vanishes on the right, so the contour can be shifted to the right without encountering any poles. Thus the contour integral equals zero.\footnote{Actually $|x|_p^{-k\gamma}$ does have a pole at $\gamma=\frac{1}{k}$, but the residue is proportional to the $p$-adic delta function $\delta_p(x)$ (see pp. 138-139 of~\cite{Gelfand:1968}), thus it does not contribute when $|x|_p > 1$. 
} 

The complex parameter $\gamma$ on the r.h.s.\ of \eqref{gammaRep} is not to be confused with the characteristic function $\gamma_p$ on the l.h.s.\ which takes a $p$-adic number as its argument. As we argue now, the complex parameter $\gamma$ has a natural interpretation as a Mellin variable. We note that the real analog of \eno{gammaRep} is the familiar integral representation of the exponential function,
\eqn{expRep}{
e^{-x} = {1 \over 2\pi i} \int_{\epsilon - i\infty}^{\epsilon+i\infty} d\gamma {\Gamma(\gamma) \over x^\gamma} \qquad \epsilon>0\,,
}
which we recognize as the statement: the inverse Mellin transform of the Euler gamma function is the exponential function.
Similarly we may think of \eno{gammaRep} (at $k=1$) as performing the inverse ($p$-adic) Mellin transform of the local zeta function $\zeta_p(\gamma)$. 

In this paper we will mostly be interested in setting $k=2$ in \eno{gammaRep}. Choosing $k=2$ is suggestive of the parallels between the Gaussian over the reals and the characteristic function of $\mathbb{Z}_{p^n}$ (and in fact also the parallels between the Euler gamma function $\Gamma(\gamma)$ and the local zeta function $\zeta_p(2\gamma)$), as summarized in the following table:
\begin{center}
\begin{tabular}{c|c}
$x \in \mathbb{R}$ & $x \in \mathbb{Q}_p$ \\ \hline\hline \\
$\displaystyle{e^{-x^2} = {1 \over 2\pi i} \int_{\epsilon - i\infty}^{\epsilon+i\infty} d\gamma\, {\Gamma(\gamma) \over x^{2\gamma}}}$ & $\displaystyle{\gamma_p(x)=\frac{2\log p}{2\pi i}\int_{\epsilon-\frac{i\pi}{2\log p}}^{\epsilon+\frac{i\pi}{2\log p}}d\gamma\,\frac{\zeta_p(2\gamma)}{|x|_p^{2\gamma}}}$ \\ \\
$\displaystyle{\Gamma(\gamma/2) = \Gamma(1) \int_{\mathbb{R}} dx\, e^{-x^2} |x|^{\gamma-1}}$ & $\displaystyle{\zeta_p(\gamma) = \zeta_p(1) \int_{\mathbb{Q}_p} dx\, \gamma_p(x) |x|_p^{\gamma-1}}$
\end{tabular}
\end{center}
where the contours in the first line are as described earlier. We will return to the identities in the second line of the table in the next subsection.

\subsection{$p$-adic integration and Schwinger parametrization}
Defining the $p$-adic units $\mathbb{U}_{p^n} \equiv \{z\in \mathbb{Q}_{p^n} \,\,:\,\,|z|_p = 1\}$, we note 
\eqn{QpUnion}{
\mathbb{Q}_{p^n}^\times = \mathbb{Q}_{p^n} \smallsetminus \{0\} = \bigsqcup_{\omega \in \mathbb{Z}} p^\omega \mathbb{U}_{p^n}\,.
}
Such a partitioning is convenient in integrating any arbitrary complex-valued function of the norm of a $p$-adic variable $x$, $f(|x|_p)$ over $\mathbb{Q}_{p^n}$, as we now describe. 

Conventionally, $p$-adic integrals are normalized by setting the Haar measure of the $p$-adic integers to $1$, namely
\eqn{intnormcond}
{\int_{\mathbb{Z}_{p^n}}dx=1\,.}
Translational invariance of the Haar measure $dx$ then dictates that
\eqn{intpomega}
{\int_{p^\omega\mathbb{U}_{p^n}}dx=\frac{p^{-n\omega}}{\zeta_p(n)} \qquad \omega \in \mathbb{Z}\,.}
Thus for an arbitrary function $f(|x|_p)$, we have
\eqn{intformula}
{\int_{\mathbb{Q}_{p^n}}dx\,f(|x|_p) = \sum_{\omega=-\infty}^{\infty} f(|p^\omega \mathbb{U}_{p^n}|_p) \int_{p^\omega\mathbb{U}_{p^n}}dx =\frac{1}{\zeta_p(n)}\sum_{\omega=-\infty}^\infty f(p^{-\omega})p^{-n\omega}\,, }
where in the second equality we used the partitioning in \eno{QpUnion} to rewrite the integral over $\mathbb{Q}_{p^n}$ as an integral over the union of open sets $p^\omega \mathbb{U}_{p^n}$, while dropping the integral over a set of measure zero. Moreover, we could pull $f(|x|_p)$ outside the integral since all elements of  $p^\omega \mathbb{U}_{p^n}$ have identical $p$-adic norm. 
As an application of this formula, one can show that
\eqn{Schwinger}
{
\frac{\zeta_p(n)}{\zeta_p(\Delta)} \int_{\mathbb{Q}_{p^n}}\frac{dS}{|S|^n_p}|S|_p^\Delta\gamma_p(xS)
=\frac{1}{|x|_p^{\Delta}}\,.
}
Equation \eqref{Schwinger} will serve for us the purpose of a $p$-adic analog to the Schwinger parameter trick over the reals, which takes the form
\eqn{ArchSchwinger}
{
\frac{1}{\Gamma(\Delta)}
\int_0^\infty \frac{dS}{S}S^{\Delta}e^{-Sx}
=\frac{1}{x^{\Delta}}\,.
}
Identities \eno{Schwinger}-\eno{ArchSchwinger} are generalizations of the identities in the second line of the table in the previous subsection.

We will also be interested in a variant of \eno{Schwinger} where the integration is over $\mathbb{Q}_p^2$, the set of $p$-adic numbers which admit a square-root in $\mathbb{Q}_p$:\footnote{The real analog of $\mathbb{Q}_p^2$ is simply $\mathbb{R}_{\geq 0}$, the set of all non-negative real numbers which was used as the integration range in \eno{ArchSchwinger}. }
\eqn{square}
{
\mathbb{Q}_p^2 \equiv \{x\in \mathbb{Q}_p \,\, : \,\, x=y^2 {\rm  \ for\ some\ } y\in\mathbb{Q}_p  \}\,.
}
We note that 
\eqn{index}
{
[\mathbb{Q}_p^\times : (\mathbb{Q}^{2}_p)^\times ]=\begin{cases}
8 &  \text{for }p=2
\\
4 &  \text{for }p>2
\end{cases}\,,
}
where $[\mathbb{Q}_p^\times : (\mathbb{Q}^{2}_p)^\times ]$ denotes the index of the multiplicative subgroup $(\mathbb{Q}^{2}_p)^\times$ in $\mathbb{Q}_p^\times$.\footnote{See e.g.\ p.\ 131ff of Ref.~\cite{Gelfand:1968}. The real analog of \eno{index} is $[\mathbb{R}^\times : \mathbb{R}^\times_{\geq 0}]=2$, where $\mathbb{R}^\times = \mathbb{R} - \{0\}$.}

From \eno{index}, together with the fact that each non-zero square in $\mathbb{Q}_p$ has precisely two square roots in $\mathbb{Q}_p$, it follows that 
\eqn{vol}
{
\int_{\mathbb{U}_p^2}dS=\frac{|2|_p}{2\,\zeta_p(1)}=\begin{cases} \displaystyle \frac{1}{4\,\zeta_2(1)} &\text{for }p=2\vspace{3mm}
\\
\displaystyle  \frac{1}{2\,\zeta_p(1)} &\text{for }p>2
\end{cases}\,.
}
This, together with a variant of \eno{intformula} for $\mathbb{Q}_p^2$ leads to the following variants of the $p$-adic Schwinger parameter trick written in \eno{Schwinger}:
\eqn{SchwingerSquare}
{
\frac{2}{|2|_p} \frac{\zeta_p(1)}{\zeta_p(2\Delta)} \int_{\mathbb{Q}^2_{p}}\frac{dS}{|S|_p}|S|_p^\Delta\gamma_p(xS)
=\frac{1}{|x|_p^{\Delta}} \quad \text{for }x\in \mathbb{Q}^2_{p}\,,
}
\eqn{SchwingerNotSquare}
{
\frac{2}{|2|_p} \frac{\zeta_p(1)}{\zeta_p(2\Delta)} \int_{p\mathbb{Q}^2_{p}}\frac{dS}{|S|_p}|S|_p^\Delta\gamma_p(xS)
=\frac{1}{|x|_p^{\Delta}} \quad \text{for }x\in p\mathbb{Q}^2_{p}\,,
}
 where
\eqn{psquare}
{
p\mathbb{Q}_p^2 \equiv \{x\in \mathbb{Q}_p \,\, : \,\, x=p y^2 {\rm  \ for\ some\ } y\in\mathbb{Q}_p  \}\,.
}

\subsection{The propagators of $p$-adic AdS/CFT}
\label{GKETC}

Finally, before we undertake the computation of Mellin amplitudes in the next section, we recall from Ref.~\cite{Gubser:2016guj} the expressions for the propagators of the bulk theory described by the free action \eqref{FreeAction} on the Bruhat-Tits tree.

The Green's function of the action gives rise to the following bulk-to-bulk propagator for a field $\phi$ of scaling dimension $\Delta$,
\eqn{bulk-to-bulk}
{
G_\Delta(z_0,z;w_0,w) = \tilde{c}_\Delta\, p^{-\Delta\,d[z_0,z;w_0,w]}  \equiv  \tilde{c}_\Delta\,\hat{G}_\Delta(z_0,z;w_0,w)\,,
}
where $\tilde{c}_\Delta$ is a normalization constant and $d[z_0,z;w_0,w]$ denotes the graph distance between the two bulk points on the tree, i.e.\ the number of edges separating the two vertices on the tree.

Taking a suitable limit of the bulk-to-bulk propagator, one can obtain the bulk-to-boundary propagator from a bulk point $(z_0,z)$ to a boundary point $x$,
\eqn{bulk-to-boundary}
{
K_\Delta(z_0,z;x) = c_{\Delta}\frac{|z_0|_p^\Delta}{|z_0,z-x|_s^{2\Delta}}  \equiv c_{\Delta}\,\hat{K}_\Delta(z_0,z;x)\,,
}
where $c_{\Delta}$ is a normalization constant and $|z_0,z-x|_s$ denotes the supremum norm, 
\eqn{}{
|z_0,z-x|_s \equiv \sup\{|z_0|_p,|z-x|_p\}\,. 
}
In this paper we adopt the following normalization convention,
\eqn{NormalizationChoice}
{
c_\Delta=\tilde{c}_\Delta=\zeta_p(2\Delta)\,.
}
This choice differs from  conventions used in Refs.~\cite{Gubser:2016guj,Gubser:2017tsi} but leads to simpler overall factors in the final expressions for Mellin amplitudes as defined by \eno{pMel}. Further, we note that when it comes to computing the Mellin amplitudes, the simple power law behavior of the propagators makes it unnecessary to pass to a ($p$-adic) embedding space formalism as is usually done in the case of real Mellin amplitudes.

\vspace{1em}

We end this section with a comment on an alternate way of writing down position space correlators such as \eno{contact} and \eno{OneExchange}, which will be especially useful in the computation of Mellin amplitudes.
Instead of starting with a discrete bulk geometry given by the Bruhat--Tits tree, one could have started with a continuum $p$-adic anti-de Sitter space given by 
\eqn{pAdS}{
p\text{AdS}_{n+1}=\mathbb{Q}_p^\times \times \mathbb{Q}_{p^n}\,,
}
where the first factor in the product represents the continuum bulk depth direction.
It turns out, owing to the ultrametricity of the $p$-adic norm, the discrete Bruhat--Tits tree ${\cal T}_{p^n}$ emerges as a course-graining of $p{\rm AdS}_{n+1}$ at AdS length scales.
This identification allows one to replace the discrete sum on the tree with a bulk integral~\cite{Gubser:2016guj},
\eqn{SumToInt}{
\sum_{(z_0,z)\in\mathcal{T}_{p^n}}f(z_0,z)=\zeta_p(1)\int_{\mathbb{Q}_p^\times}{dz_0 \over |z_0|_p^{n+1}} \int_{Q_{p^n}} dz\, f(z_0,z)\,,
}
for any function $f(z_0,z)$  which takes a constant value over each ball $B(z_0,z) \equiv z_0 \mathbb{U}_p \times (z+z_0 \mathbb{Z}_{p^n})$. The ball $B(z_0,z)$ corresponds precisely to the set of points in $p{\rm AdS}_{n+1}$ which are up to a unit AdS length separated from $(z_0,z)$ as measured using a chordal distance function.
Roughly, equation \eqref{SumToInt} can be understood as follows: Each bulk point $(z_0,z)$ is identified with a subset of boundary points, and $z$ is one representative from this set. But rather than picking an arbitrary representative, we can integrate $z$ over the whole subset, provided we also include a factor of $|z_0|_p^{-n}$ to compensate for the overcounting. As for $z_0$, one could have restricted this variable to run over all values of $p^{\omega }$ with $\omega\in \mathbb{Z}$, but instead the right-hand side of equation \eqref{SumToInt} integrates $z_0$ over all of $\mathbb{Q}_p^\times$ and compensates for the overcounting with a factor of $\zeta_p(1)/|z_0|_p$ in the integrand.  

It is easily checked that the bulk-to-bulk and bulk-to-boundary propagators written above are examples of functions $f(z_0,z)$ which satisfy \eno{SumToInt}. Thus  we may rewrite, for instance the position space contact amplitude \eno{contact}, as
\eqn{contactRewrite}{
{\cal A}^{\rm contact}(x_i) = \sum_{(z_0,z) \in {\cal T}_{p^n}} \prod_{i} K_{\Delta_i}(z_0,z;x_i)
 = \zeta_p(1)\int_{p{\rm AdS}_{n+1}}{dz_0\, dz \over |z_0|_p^{n+1}}  \prod_{i} K_{\Delta_i}(z_0,z;x_i)\,,
}
which now looks similar to the usual prescription for computing correlators in the standard AdS/CFT correspondence.

\section{$p$-adic Mellin Amplitudes}
\label{contactSection}

In this section we build on the previously discussed tools and techniques to compute the $p$-adic Mellin amplitude of the $\mathcal{N}$-point contact diagram for arbitrary $\mathcal{N}$, followed by arbitrary-point amplitudes for bulk diagrams with one, two and three internal lines.

\subsection{$\mathcal{N}$-point contact diagram}
\label{CONTACT}

The first Mellin amplitude we will compute is the Mellin amplitude for the contact diagram for ${\cal N}$ external scalar insertions. We guessed in  section \ref{EXAMPLE4PT} that this amplitude (for ${\cal N}=4$) is a constant, given by \eqref{padicContact}, and used that to reproduce the position space amplitude. In this section we explicitly derive this result for arbitrary ${\cal N}$ by re-expressing the position-space contact amplitude \eqref{contact} in the form \eqref{pMel}, from which we can simply read off the Mellin amplitude $\mathcal{M}$.

Substituting \eqref{bulk-to-boundary} with the normalization \eqref{NormalizationChoice} for the bulk-to-boundary propagators in \eqref{contact}, we have 
\eqn{contactAgain}{
\mathcal{A}^{\text{con}}(x_i)
=
\sum_{(z_0,z)\in \mathcal{T}_{p^n}} \prod_{i=1}^\mathcal{N}\zeta_p(2\Delta_i)\frac{|z_0|_p^{\Delta_i}}{|z_0,z-x_i|_s^{2\Delta_i}}\,.
}
Using \eno{SumToInt} to convert the discrete summation to a continuum integral, 
we obtain 
\eqn{}
{
\mathcal{A}^{\text{con}}(x_i)
=
\zeta(1)\int_{\mathbb{Q}_p} \frac{dz_0}{|z_0|_p} |z_0|_p^{\sum_i\Delta_i-n}\int_{\mathbb{Q}_{p^n}}dz\prod_{i=1}^\mathcal{N}\frac{\zeta_p(2\Delta_i)}{|z_0,z-x_i|_s^{2\Delta_i}}\,,
}
where the domain of the $z_0$ integral has been extended by a measure zero set (recall that $\mathbb{Q}_p = \mathbb{Q}_p ^\times \sqcup \{0\}$).

At this point it is useful to invoke the $p$-adic Schwinger-parametrization given in \eqref{SchwingerSquare} as well as the factorization property \eqref{GammaFactorization} to re-express $\mathcal{A}^{\text{con}}(x_i)$ as
\eqn{z0z}{
&\zeta(1) 
\prod_{i=1}^\mathcal{N}\bigg(\frac{2\zeta_p(1)}{|2|_p}\int_{\mathbb{Q}^2_p}\frac{dS_i}{|S_i|_p}|S_i|_p^{\Delta_i}\bigg)
\int_{\mathbb{Q}_p}\frac{dz_0}{|z_0|_p} |z_0|_p^{\sum_{i}\Delta_i-n}\int_{\mathbb{Q}_{p^n}}dz
\prod_{i=1}^\mathcal{N}\bigg(\gamma_p(S_iz_0^2)\gamma_p(S_i(z-x_i)^2)\bigg).
}
Let $m$ be an index such that $|S_m|_p = \text{sup}(|S_1|_p,...,|S_\mathcal{N}|_p)$. Then the $z_0$ integral above can immediately be carried out to give,
\eqn{}
{
\int_{\mathbb{Q}_p} \frac{dz_0}{|z_0|_p} |z_0|_p^{\sum_{i}\Delta_i-n} \gamma_p(S_mz_0^2)
=\frac{\zeta_p(\sum_{i}\Delta_i-n)}{\zeta(1)|S_m|_p^{\sum_{i}\Delta_i/2-n/2}}\,.
}
Turning to the $z$ integral, we first shift the variable $z$ by $x_m$. Note that a factor of $\gamma_p(S_m z^2)$ forces $S_i z^2$ to be a $p$-adic integer for all $i =1,...,\mathcal{N}$ on the support of the integrand, which implies that $\gamma_p(S_i (z-x_{im})^2) = \gamma_p(S_i x_{im}^2)$. So translating $z$ by $x_m$, the only non-trivial $z$-dependence in \eno{z0z} comes from the characteristic function $\gamma_p(S_m z^2)$, and this leads to an $x$-independent $z$-integral,
\eqn{}
{
\int_{\mathbb{Q}_{p^n}}dz\,\gamma_p(S_mz^2)=\frac{1}{|S_m|_p^{n/2}}\,,
}
which can be obtained from the Schwinger parameter identity \eno{Schwinger}.
Combining the previous two results, we get 
\eqn{Acon1}{
\mathcal{A}^{\text{con}}(x_i)=
\zeta_p\left(\sum_{i}\Delta_i-n\right)
\prod_{i=1}^\mathcal{N}\bigg(\frac{2\zeta_p(1)}{|2|_p}\int_{\mathbb{Q}^2_p}\frac{dS_i}{|S_i|_p}|S_i|_p^{\Delta_i}\bigg)\frac{1}{|S_m|_p^{\sum_{i}\Delta_i/2}}\prod_{i\neq m}\bigg(\gamma_p(S_i x_{im}^2)\bigg)\,.
}
We now rewrite factors of the characteristic function $\gamma_p(S_i x_{im}^2)$ as $\gamma_p(\frac{S_iS_m}{S_m} x_{im}^2)$. Since $x_{ij}=x_{im}+x_{mj}$, it follows from the ultra-metricity of the $p$-adic norm that $|x_{ij}|_p\leq \text{max}(|x_{im}|_p,|x_{mj}|_p)$. Furthermore, $|S_iS_j|_p \leq |S_iS_m|_p, |S_jS_m|_p$. It then follows that $\gamma_p(\frac{S_iS_j}{S_m}x_{ij}^2)$ is equal to unity on the support of $\gamma_p(\frac{S_iS_m}{S_m}x_{im}^2)\gamma_p(\frac{S_jS_m}{S_m}x_{jm}^2)$. We conclude that
\eqn{prodgamma1}
{
\prod_{i\neq m}\gamma_p(S_i x_{im}^2)=\prod_{1\leq i < j\leq \mathcal{N}}\gamma_p\bigg(\frac{S_iS_j}{S_m}x_{ij}^2\bigg)\,.
}
At this point we introduce new variables $s_i$, defined to be
\eqn{varChange}{
s_i \equiv \sqrt{|S_m|_p}\,S_i
\hspace{5mm} \text{for }i\neq m\,, 
\hspace{10mm} s_m\equiv \sqrt{S_m}\,.
}
We can take square-roots in \eno{varChange} since as is clear from \eno{z0z}, $S_i \in \mathbb{Q}_p^2$ for all $i$ and thus admit square-roots in $\mathbb{Q}_p$ -- we will specify precisely {\it which} square-root did we mean in \eno{varChange} shortly. 
Just like the familiar change of variables over the real or complex fields, one picks up a Jacobian factor. In this case we pick up a factor of $|2s_m^{\mathcal{N}}|_p$.
It is worth emphasizing that the change of variables \eno{varChange} makes explicit reference to the index $m$, defined below \eno{z0z}. 
Thus the value that $m$ takes is $S_i$ dependent, and so varies in the domain of integration over $S_i$.
 Therefore, the change of variables is well-defined only if we partition the original integration domain into subsets each admitting a fixed value of $m$, and find new variables $s_i$ for each such sub-domain.
This partitioning is somewhat concealed by the notation adopted here, but the change of variables remains perfectly valid nonetheless. We now describe the domain of integration in the new variables $s_i$.

We note that the domain of $s_m$ is ``half'' the $p$-adic numbers, in the sense that it is all the $p$-adic numbers with distinct squares. 
Since $S_m \in \mathbb{Q}_p^2$ has precisely two square-roots, say $x,y$ such that $x^2=y^2=S_m$, let us specify which square-root goes in \eno{varChange}. First note that, $y=-x$. Now the $p$-adic number $x$ has a unique power series expansion $x = p^v \hat{x}$, where $v \in \mathbb{Z}$ and $\hat{x} \in \mathbb{U}_p$, i.e.\ $\hat{x} = x_0 + x_1 p + x_2 p^2 + \cdots$ with $x_0 \in \{1,\ldots, p-1\}$, and similarly for $y$.  So $y=-x \Rightarrow y_0 = p-x_0$, which implies that for $p>2$, $y_0$ is a square mod $p$ iff $x_0$ is not. So for $p>2$, we prescribe that the square-root in \eno{varChange} is the one whose units digit is a square mod $p$. Let's say this square-root is $x$, which implies in fact $\hat{x} \in \mathbb{U}_p^2$. Then  $s_m=x = p^v \hat{x}$ either belongs to $\mathbb{Q}_p^2$ (for even $v$) or $p\mathbb{Q}_p^2$ (for odd $v$), as we sweep across the domain of $S_m$. This is what we meant by ``half'' the $p$-adic numbers.

If we restrict $s_m$ to the domain $\mathbb{Q}_p^2 \cup p \mathbb{Q}_p^2$ for $p=2$, we must also multiply by an overall factor of two, in light of equation \eqref{index}. The upshot is that we can  take the domain of $s_m$ to be $\mathbb{Q}_p^2 \cup p\mathbb{Q}_p^2$ provided we introduce a factor of $1/|2|_p$, which exactly cancels the factor of $|2|_p$ that we pick up from the Jacobian. 
Now note that it follows from \eqref{varChange} that if $s_m \in \mathbb{Q}_p^2$, then $s_i \in \mathbb{Q}_p^2$ for all $i$, and if $s_m \in p\mathbb{Q}_p^2$, then $s_i \in p\mathbb{Q}_p^2$ for all $i$. 
Plugging in the new variables in \eno{Acon1}-\eno{prodgamma1}, we obtain an expression for the contact amplitude in the new variables, where all reference to the index $m$ has vanished entirely,
\eqn{preMellin}{
\mathcal{A}^{\text{con}}=\sum_{a\in\{1,p\}}
\zeta_p\big(\sum_{i}\Delta_i-n\big)
\prod_{i}\bigg(\frac{2\zeta_p(1)}{|2|_p}\int_{a\mathbb{Q}^2_p}\frac{ds_i}{|s_i|_p}|s_i|_p^{\Delta_i}\bigg)\prod_{i < j}\gamma_p\bigg(s_is_jx_{ij}^2  \bigg)\,.
}

Now we will invoke the Mellin representation  of the characteristic function given in \eqref{gammaRep}. Similarly to the Archimedean case where the Mellin variables are subject to $\mathcal{N}$ constraints that can be interpreted as momentum conservation in an auxiliary space, we apply \eqref{gammaRep} to only $\mathcal{N}(\mathcal{N}-3)/2$ of the $\mathcal{N}(\mathcal{N}-1)/2$ factors of $\gamma_p(s_is_jx_{ij}^2)$ in \eno{preMellin}. For concreteness, we pick these factors to be the ones for which $i,j\geq 2$ except $(i,j)=(2,3)$, though any other choice will work just as well. Doing this, we get 
\eqn{}{
\mathcal{A}^{\text{con}} &=
\zeta_p\big(\sum_{i}\Delta_i-n\big)\prod_{\substack{2\leq i < j \leq \mathcal{N} \\ (i,j)\neq (2,3)}} \bigg(\frac{\log p}{\pi i}\int d\gamma_{ij}
 \frac{\zeta_p(2\gamma_{ij})}{|x_{ij}|_p^{2\gamma_{ij}}}\bigg)
\frac{2^3\zeta_p(1)^3}{|2|_p^3} \cr
 & \quad \times 
\sum_{a \in \{1, p\}} \int_{a\mathbb{Q}_p^2} \frac{ds_1}{|s_1|_p}\frac{ds_2}{|s_2|_p}\frac{ds_3}{|s_3|_p}|s_1|_p^{\Delta_1}|s_2|_p^{\Delta_2-\sum_{i=4}^\mathcal{N}\gamma_{2i}}|s_3|_p^{\Delta_3-\sum_{i=4}^\mathcal{N}\gamma_{3i}}\gamma_p\bigg(s_1s_2x_{12}^2 \bigg)\cr
& \quad \times
\gamma_p\bigg(s_1s_3x_{13}^2 \bigg)\gamma_p\bigg(s_2s_3x_{23}^2 \bigg)
\prod_{i=4}^\mathcal{N}\left[\frac{2\zeta_p(1)}{|2|_p}\int_{a\mathbb{Q}_p^2} \frac{ds_i}{|s_i|_p} \Big(|s_i|_p\Big)^{\Delta_i-\sum_{\substack{j=2\\j\neq i}}^\mathcal{N}\gamma_{ij}} \gamma_p\bigg(s_1s_ix_{1i}^2 \bigg)\right].
}
The integrals over $s_i$ for $i=4,...,\mathcal{N}$ factor out and can be carried out directly using equations \eqref{SchwingerSquare} and \eqref{SchwingerNotSquare}. If we introduce the following definitions, 
\eqn{}
{
 \gamma_{23} \equiv \Delta_3-\gamma_{13}-\sum_{j=4}^\mathcal{N}\gamma_{3j} \qquad  \gamma_{1i} \equiv \Delta_i-\sum_{\substack{j=2\\j\neq i}}^\mathcal{N}\gamma_{ij} \qquad  i=2,...,\mathcal{N}\,,
}
which are consistent with the constraints \eno{MellinVarConstraints} obeyed by the Mellin variables of an ${\cal N}$-point Mellin amplitude, we can rewrite
\eqn{nearlyDone}{
\mathcal{A}^{\text{con}} &=
\sum_{a \in \{1, p\}}\prod_{\substack{2\leq i < j \leq \mathcal{N} \\ (i,j)\neq (2,3)}} \bigg(\frac{\log p}{\pi i}\int d\gamma_{ij}
 \frac{\zeta_p(2\gamma_{ij})}{|x_{ij}|_p^{2\gamma_{ij}}}\bigg)
\prod_{i=4}^\mathcal{N}\bigg(\zeta_p(2\gamma_{1i})\frac{1}{|x_{1i}|_p^{2\gamma_{1i}}}\bigg)
\frac{2^3\zeta_p(1)^3}{|2|_p^3}\zeta_p\big(\sum_{i}\Delta_i-n\big)
 \cr
 & \times
\int_{a\mathbb{Q}_p^2} \frac{ds_1}{|s_1|_p}\frac{ds_2}{|s_2|_p}\frac{ds_3}{|s_3|_p}|s_1|_p^{\gamma_{12}+\gamma_{23}}|s_2|_p^{\gamma_{12}+\gamma_{13}}|s_3|_p^{\gamma_{13}+\gamma_{23}}\gamma_p\bigg(s_1s_2x_{12}^2 \bigg)\gamma_p\bigg(s_1s_3x_{13}^2 \bigg)\gamma_p\bigg(s_2s_3x_{23}^2 \bigg).
}
Here it is helpful to do one more change of variables,
\eqn{}{
 T_1 \equiv s_2s_3, \hspace{10mm} T_2 \equiv s_1s_3, \hspace{10mm} T_3 \equiv s_1s_2\,.
}
Since we are requiring that all the $s_i$ belong to either $\mathbb{Q}^2_p$ or $p\mathbb{Q}^2_p$, it follows that the $T_i$ are squares in $\mathbb{Q}_p$. Furthermore, integrating each of $T_1$, $T_2$, and $T_3$ over all of $\mathbb{Q}^2_p$ will exactly reproduce the integral of all the $s_i$ over $\mathbb{Q}_p^2$ plus the integral of all the $s_i$ over $p\mathbb{Q}_p^2$. We can therefore lump the $a=1$ and the $a=p$ terms in \eqref{nearlyDone} together by changing to the $T_i$ variables. The $T_i$ integrals can then be carried out using \eqref{SchwingerSquare} to give
\eqn{}{
\mathcal{A}^{\text{con}}=
\prod_{\substack{2\leq i < j \leq \mathcal{N} \\ (i,j)\neq (2,3)}} \bigg(\frac{\log p}{\pi i}\int d\gamma_{ij}
 \frac{\zeta_p(2\gamma_{ij})}{|x_{ij}|_p^{2\gamma_{ij}}}\bigg)
\prod_{i=2}^\mathcal{N}\bigg(\frac{\zeta_p(2\gamma_{1i})}{|x_{1i}|_p^{2\gamma_{1i}}}\bigg)
\frac{\zeta_p(\sum_{i}\Delta_i-n)
\zeta_p(2\gamma_{23})}{|x_{23}|_p^{2\gamma_{23}}}\,.
}

This form of the contact diagram reflects the arbitrary choice made in picking which characteristic functions to express in the Mellin representation \eqref{gammaRep}. To re-write the diagram in a more symmetric fashion, we define\footnote{The definition \eno{GammaMeasure} is precisely equivalent to the definition given earlier in \eno{pMellinMeasure}.}
\eqn{GammaMeasure}{
[d\gamma]\equiv\left(\frac{\log p}{\pi i}\right)^{\frac{\mathcal{N}(\mathcal{N}-3)}{2}}\left[\prod_{1\leq i < j \leq \mathcal{N}}d\gamma_{ij}\right]\left[\prod_{i=1}^\mathcal{N}\delta\big(\sum_{j=1}^\mathcal{N}\gamma_{ij}\big)\right], \hspace{10mm} \gamma_{ij}=\gamma_{ji}, \hspace{5mm}\gamma_{ii}=-\Delta_i\,,
}
which immediately gives
\eqn{contactResult}{
\mathcal{A}^{\text{con}}=
\zeta_p\big(\sum_{i}\Delta_i-n\big)\int [d\gamma] \prod_{1\leq i<j\leq {\cal N}}\frac{\zeta_p(2\gamma_{ij})}{|x_{ij}|_p^{2\gamma_{ij}}}\,.
}
We conclude that the Mellin amplitude for the ${\cal N}$-point contact diagram for external scalar insertions is
\eqn{contactFinal}{
\mathcal{M}^{\text{con}}=
\zeta_p\big(\sum_i \Delta_i-n\big)\,.
}
Readers familiar with the corresponding calculation of the contact amplitude over the reals may be able to appreciate the similarity with multiple intermediate steps in this derivation.

\vspace{1em}
It is worth remarking that from comparing \eqref{preMellin} with \eqref{contactResult}, one obtains the $p$-adic analog of the Symanzik star integration formula~\cite{Symanzik:1972wj} (see also appendix B of Ref.~\cite{Paulos:2011ie}),
\eqn{SymanzikStar}{
\int [d\gamma] \prod_{1\leq i<j\leq {\cal N}}\frac{\zeta_p(2\gamma_{ij})}{|x_{ij}|_p^{2\gamma_{ij}}} = 
   \sum_{a\in \{1,p\}}  \prod_{i=1}^{\cal N} \left( \frac{2\zeta_p(1)}{|2|_p} \int_{a\mathbb{Q}^2_p} \frac{ds_i}{|s_i|_p}|s_i|_p^{\Delta_i}\right) \prod_{1\leq i<j\leq {\cal N}}\gamma_p\bigg(s_is_jx_{ij}^2  \bigg)\,.
}

\subsection{The split representation of the bulk-to-bulk propagator}
\label{HARMEXPAND}

In computing the Mellin amplitudes for exchange diagrams, it will be useful to re-express the $p$-adic bulk-to-bulk propagator in its split representation as given in \eno{HarmExp}, in much the same way as the spectral decomposition of the bulk-to-bulk propagator \eno{HarmExpReal} is a useful first step when computing real Mellin amplitudes~\cite{Penedones:2010ue}. We rewrite \eno{HarmExp} as follows,
\eqn{HarmExpAgain}
{
G_{\Delta}(z_0,z;w_0,w)=
\frac{1}{2}
\frac{\log p}{2\pi i} &\int_{-\frac{i\pi}{\log p}}^{\frac{i\pi}{\log p}}\, dc\,\frac{\zeta_p\big(\Delta-\frac{n}{2}+c\big)\zeta_p\big(\Delta-\frac{n}{2}-c\big)}{\zeta_p(2c)\zeta_p(-2c)}
\cr
\times &
\int_{\mathbb{Q}_{p^n}}dx\,K_{\frac{n}{2}-c}(z_0,z;x)K_{\frac{n}{2}+c}(w_0,w;x)\,.
}
In this subsection we will prove this identity. 

One starts by computing the following integral, 
\eqn{KaKb}{
\int_{\mathbb{Q}_{p^n}}dx\,\hat{K}_a(z_0,z;x)\hat{K}_b(w_0,w;x)\,,
}
where we point out that the bulk-to-boundary propagators above are the unnormalized propagators defined in \eno{bulk-to-boundary}.
We plug in the explicit form of the bulk-to-boundary propagator \eqref{bulk-to-boundary} and then use the Schwinger parameter trick \eqref{Schwinger} to re-express all powers, to get the following equivalent form for the integral \eqref{KaKb},
\eqn{KaKbInt}
{
&\frac{\zeta_p(n)^2|z_0|_p^a|w_0|_p^b}{\zeta_p(2a)\zeta_p(2b)}
\cr
&\times\int_{\mathbb{Q}_{p^n}}\!\!
dx\int_{\mathbb{Q}_{p^n}}\!\! \frac{dS_a}{|S_a|_p^n}\int_{\mathbb{Q}_{p^n}}\!\! \frac{dS_b}{|S_b|_p^n}\, |S_a|_p^{2a}|S_b|_p^{2b}
\gamma_p(S_az_0)\gamma_p\big(S_a(z-x)\big)
\gamma_p(S_bw_0)\gamma_p\big(S_b(w-x)\big).
}
The integrals over $x$, $S_a$, and $S_b$ can be evaluated by splitting the integration domain into the region where $|S_a|_p \geq |S_b|_p$ (obtained by introducing a factor of $\gamma_p(S_b/S_a)$ in the integrand)  and the region where $|S_b|_p \geq |S_a|_p$, and finally subtracting off the doubly-counted region where $|S_a|_p=|S_b|_p$. For each of these three parts, the $x$ integral can be carried out immediately using \eqref{GammaFactorization} and \eqref{Schwinger}. An intermediate result that is useful for evaluating the remaining $S_a$ and $S_b$ integrals is
\eqn{}
{
\int_{\mathbb{Q}_{p^n}}\frac{dS}{|S|^n}|S|_p^{\Delta}\gamma_p\bigg(\frac{S}{A}\bigg)\gamma_p\bigg(\frac{B}{S}\bigg)=\bigg[\frac{\zeta_p(-\Delta)}{\zeta_p(n)}|B|_p^\Delta+\frac{\zeta_p(\Delta)}{\zeta_p(n)}|A|_p^\Delta\bigg]\gamma_p\left(\frac{B}{A}\right).
}
After some work, \eqref{KaKbInt} evaluates to
\eqn{}{
&
\bigg[\zeta_p(n-2b)+\zeta_p(n-2a)-1\bigg]\frac{\zeta_p(2a+2b-n)}{\zeta_p(2a)\zeta_p(2b)}\frac{|z_0|_p^a|w_0|_p^b}{|z_0,w_0,z-w|_s^{2a+2b-n}}
\cr
+ &
 \frac{\zeta_p(2b-n)\zeta_p(2a)}{\zeta_p(2a)\zeta_p(2b)}\frac{|w_0|_p^{n-b}|z_0|_p^a}{|z_0,w_0,z-w|_s^{2a}}+
\frac{\zeta_p(2a-n)\zeta_p(2b)}{\zeta_p(2a)\zeta_p(2b)}\frac{|z_0|_p^{n-a}|w_0|_p^b}{|z_0,w_0,z-w|_s^{2b}}\,.
}
Setting $a=\frac{n}{2}-c$ and $b=\frac{n}{2}+c$, and restoring the normalizations of the bulk-to-boundary propagators using \eno{bulk-to-boundary}, we find that
\eqn{}{
&\int_{\mathbb{Q}_{p^n}}dx\,K_{\frac{n}{2}-c}(z_0,z;x)K_{\frac{n}{2}+c}(w_0,w;x)
\cr
& =
\zeta_p(2c)\zeta_p(n-2c)\frac{|w_0|_p^{\frac{n}{2}-c}|z_0|_p^{\frac{n}{2}-c}}{|z_0,w_0,z-w|_s^{n-2c}}+
\zeta_p(-2c)\zeta_p(n+2c)\frac{|z_0|_p^{\frac{n}{2}+c}|w_0|_p^{\frac{n}{2}+c}}{|z_0,w_0,z-w|_s^{n+2c}}\,.
}
Using this result we can proceed to calculate the right-hand side of \eno{HarmExpAgain}. It is necessary, however, to distinguish between the cases where the bulk points $(z_0,z)$ and $(w_0,w)$ are coincident and non-coincident. 

When $(z_0,z)=(w_0,w)$, the r.h.s.\ of \eno{HarmExpAgain} reduces to
\eqn{}
{
\frac{1}{2}\frac{\zeta_p(n)}{\zeta_p(2n)}\frac{\log p}{2\pi i} \int_{-\frac{i\pi}{\log p}}^{\frac{i\pi}{\log p}}\, dc\,\frac{\zeta_p\big(\Delta-\frac{n}{2}-c\big)\zeta_p\big(\Delta-\frac{n}{2}+c\big)\zeta_p(n-2c)\zeta_p(n+2c)}{\zeta_p(2c)\zeta_p(-2c)}\,.
}
The contour can be closed in either direction. One must either sum up the residues at the poles situated at $c=\Delta-\frac{n}{2}$, $c=\frac{n}{2}$ and $c=\frac{n}{2}+\frac{i\pi}{\log p}$, or the residues at the poles situated at minus these locations. The result is simply $\zeta_p(2\Delta)$, which exactly equals $G_{\Delta}(z_0,z;w_0,w)$ for coincident points $(z_0,z)=(w_0,w)$ (see \eno{bulk-to-bulk}). This verifies the split representation \eqref{HarmExpAgain} for coincident points.

If $(z_0,z) \neq (w_0,w)$, then $|z_0,w_0,z-w|_s=|z_0-w_0,z-w|_s$ and the r.h.s.\ of \eno{HarmExpAgain} is equal to\footnote{Here we used the following identity between two distinct bulk points $(z_0,z)$ and $(w_0,w)$ on the Bruhat--Tits tree~\cite{Gubser:2016guj},
\eqn{}{
{\left|z_0 w_0\right|_p \over |z_0-w_0,z-w|_s^2} = p^{-d[z_0,z;w_0,w]}\,.
}
}
\eqn{}{
\frac{1}{2}\frac{\log p}{2\pi i} 
&
\int_{-\frac{i\pi}{\log p}}^{\frac{i\pi}{\log p}}\, dc\,\zeta_p\big(\Delta-\frac{n}{2}-c\big)\zeta_p\big(\Delta-\frac{n}{2}+c\big)
\cr
\times \left[\frac{\zeta_p(n-2c)}{\zeta_p(-2c)} \right.
& 
\left. \hat{G}_{\frac{n}{2}-c}(z_0,z;w_0,w)+\frac{\zeta_p(n+2c)}{\zeta_p(2c)}\hat{G}_{\frac{n}{2}+c}(z_0,z;w_0,w)\right].
}
The contour must be closed on the left for the first term and on the right for the second term since the bulk-to-bulk propagator between two non-coincident points tends to zero as the scaling dimension tends to zero. Note that we are assuming $\Delta > \frac{n}{2}$. The first term then picks up the residue from the pole at $c=-(\Delta-\frac{n}{2})$, and the second term picks up the residue from the pole at $c=\Delta-\frac{n}{2}$. The two terms yield the same result, adding up to give
\eqn{}{
\zeta_p(2\Delta)\hat{G}_{\Delta}(z_0,z;w_0,w)=G_{\Delta}(z_0,z;w_0,w)\,.
}
This completes the proof of the split representation \eqref{HarmExpAgain}.

\subsection{Exchange diagrams}
\label{singleSection}

With the split representation in hand, we are ready to evaluate exchange diagrams. Consider the diagram:
\eqn{TheExchangeDiagram}{
\begin{matrix}
\includegraphics[height=14ex]{figures/single.pdf}
 \end{matrix}\,.
}
Generally, the difficulty in computing Mellin amplitudes increases with number of internal lines but is insensitive to the number of external legs and the dimensions of operators, so we may as well consider the general case where an unspecified number of external insertions at the boundary,  carrying generic scaling dimensions that are labeled by a dummy index $i_L$, are incident on the internal leg on the left, while the external legs to the right carry the dummy index $i_R$. We denote the scaling dimension of the scalar operator exchanged along the internal line by $\Delta_A$. Then, the position space amplitude is given by
\eqn{}{
\mathcal{A}^{\text{exc}}=
\sum_{(z^L_0,z^L),(z^R_0,z^R)\in \mathcal{T}_{p^n}}\prod_{i_L}\bigg(K_{\Delta_{i_L}}(z^L_0,z^L;x_{i_L})\bigg)
\prod_{i_R}\bigg(K_{\Delta_{i_R}}(z^R_0,z^R;x_{i_R})\bigg)G_{\Delta_A}(z^L,z^L_0;z^R,z^R_0)\,.
}
We re-express $G_{\Delta_A}$ in its split representation \eqref{HarmExpAgain}, so that the integrand takes the form of a product of two contact diagrams, to which we may apply the result for contact amplitudes from above, \eqref{preMellin} to get
\eqn{ExcContour}{
\mathcal{A}^{\text{exc}}=
&
\frac{1}{2}
\frac{\log p}{2\pi i} \int_{-\frac{i\pi}{\log p}}^{\frac{i\pi}{\log p}}\, dc\,\frac{\zeta_p\big(\Delta_A-\frac{n}{2}+c\big)\zeta_p\big(\Delta_A-\frac{n}{2}-c\big)}{\zeta_p(2c)\zeta_p(-2c)}
\tilde{\mathcal{A}}^{\text{exc}}\,,
}
where $\tilde{\mathcal{A}}^{\text{exc}}$, which following Ref.~\cite{Yuan:2017vgp} we will refer to as the (position space) ``pre-amplitude'', is given by
\eqn{ExchangeInitial}{
\tilde{\mathcal{A}}^{\text{exc}} &= \int_{\mathbb{Q}_{p^n}}  dx \cr 
& \times \!\!\!
\sum_{a_L\in\{1,p\}}\zeta_p\big(\sum \Delta_{i_L}-\frac{n}{2}-c\big)
\prod_{i_L}\bigg(\frac{2\zeta_p(1)}{|2|_p}\int_{a_L\mathbb{Q}^2_p}\frac{ds_{i_L}}{|s_L|_p}|s_{i_L}|_p^{\Delta_{i_L}}\bigg)\prod_{i_L < j_L}\gamma_p\bigg(s_{i_L}s_{j_L}x_{i_Lj_L}^2  \bigg)
\cr
& \times \!\!\!
\sum_{a_R\in\{1,p\}}
\zeta_p\big(\sum \Delta_{i_R}-\frac{n}{2}+c\big)
\prod_{i_R}\bigg(\frac{2\zeta_p(1)}{|2|_p}\int_{a_R\mathbb{Q}^2_p}\frac{ds_{i_R}}{|s_{i_R}|_p}|s_{i_R}|_p^{\Delta_{i_R}-1}\bigg)\prod_{ i_R < j_R }\gamma_p\bigg(s_{i_R}s_{j_R}x_{i_Rj_R}^2  \bigg)
\cr
& \times 
\frac{2\zeta_p(1)}{|2|_p}\int_{a_L\mathbb{Q}^2_p} \frac{dt_L}{|t_L|_p}|t_L|_p^{\frac{n}{2}-c} \prod_{i_L} \gamma_p\bigg(t_Ls_{i_L}(x_{i_L}-x)^2\bigg)\cr 
&  \times \frac{2\zeta_p(1)}{|2|_p} \int_{a_R\mathbb{Q}^2_p} \frac{dt_R}{|t_R|_p}|t'|_p^{\frac{n}{2}+c} \prod_{i_R} \gamma_p\bigg(t_Rs_{i_R}(x_{i_R}-x)^2\bigg)\,.
}
Note that we are adopting a notational convention where the indices $i_L,j_L$ represent external legs on the left side of the exchange diagram \eno{TheExchangeDiagram}, while indices $i_R,j_R$ represent represent external legs on the right side. We sometimes omit explicitly specifying the domain a sum or product is taken over when it should be clear from the summand; e.g. the sum $\sum\Delta_{i_L}$ is to be understood as the sum over all the external scaling dimensions of the external legs that lie to the left of the internal leg.

On changing variables in \eno{ExchangeInitial} by introducing $S_{i_L} \equiv \frac{s_{i_L}}{t_{L}}$ and
$S_{i_R} \equiv \frac{s_{i_R}}{t_{R}}$, one is left with an $x$-integral that can be evaluated using the same reasoning as the $z$ integral in \eqref{z0z}. Then $\tilde{\mathcal{A}}^{\text{exc}}$ reduces to:
\eqn{}
{
& \quad 
\zeta_p\big(\sum \Delta_{i_L}-\frac{n}{2}-c\big)
\prod_{i_L}\bigg(\frac{2\zeta_p(1)}{|2|_p}\int_{\mathbb{Q}^2_p}\frac{dS_{i_L}}{|S_{i_L}|_p}|S_{i_L}|_p^{\Delta_{i_L}}\bigg)\prod_{ i_L < j_L}\gamma_p\bigg(\frac{S_{i_L}S_{j_L}}{t_L^2}x_{i_Lj_L}^2  \bigg)\prod_{i_L} \gamma_p\bigg(S_{i_L}x_{i_Lm}^2\bigg)
\cr
 & \times
\zeta_p\big(\sum \Delta_{i_R}-\frac{n}{2}+c\big)
\prod_{i_R}\bigg(\frac{2\zeta_p(1)}{|2|_p}\int_{\mathbb{Q}^2_p}\frac{dS_{i_R}}{|S_{i_R}|_p}|S_{i_R}|_p^{\Delta_{i_R}}\bigg)\prod_{i_R < j_R}\gamma_p\bigg(\frac{S_{i_R}S_{j_R}}{t_R^2}x_{i_Rj_R}^2  \bigg)\prod_{i_R} \gamma_p\bigg(S_{i_R}x_{i_Rm}^2\bigg)
\cr
& \times 
\frac{2^2\zeta_p(1)^2}{|2|_p^2}\sum_{a_L,a_R}\int_{a_L\mathbb{Q}^2_p} \frac{dt_L}{|t_L|_p}
|t_L|_p^{\frac{n}{2}-\sum\Delta_{i_L}-c}
\int_{a_R\mathbb{Q}^2_p}\frac{dt_R}{|t_R|_p}
|t_R|_p^{\frac{n}{2}-\sum\Delta_{i_R}+c}|S_m|_p^{-\frac{n}{2}}
\,,
}
where, as in section \ref{contactSection}, $m$ is an index such that $|S_m|_p=\text{sup}|S_i|_p$ where $i$ runs over all values that $i_L$ and $i_R$ take. 
By changing variables so that $t_L\rightarrow S_mt_L$ and $t_R\rightarrow S_mt_R$ and then changing the variables $S_i$ to new variables $s_i$ analogously to the change of variables \eqref{varChange},  one finds that $\tilde{\mathcal{A}}^{\text{exc}}$ is equal to
\eqn{preMellinEx}{
&
\zeta_p\big(\sum \Delta_{i_L}-\frac{n}{2}-c\big)\zeta_p\big(\sum \Delta_{i_R}-\frac{n}{2}+c\big)\int_{\mathbb{Q}_{p}} \frac{dt_L}{|t_L|_p}\int_{\mathbb{Q}_{p}}\frac{dt_R}{|t_R|_p}|t_L|_p^{\frac{n}{2}-\sum\Delta_{i_L}-c} 
|t_R|^{\frac{n}{2}-\sum\Delta_{i_R}+c}
\cr
& \times 
\zeta_p(1)^2\sum_{a\in\{1,p\}}\prod_{i_L}\bigg(\frac{2\zeta_p(1)}{|2|_p}\int_{a\mathbb{Q}^2_p}\frac{ds_{i_L}}{|s_{i_L}|_p}|s_{i_L}|_p^{\Delta_{i_L}}\bigg)
\prod_{i_R}\bigg(\frac{2\zeta_p(1)}{|2|}\int_{a\mathbb{Q}^2_p}\frac{ds_{i_R}}{|s_{i_R}|_p}|s_{i_R}|_p^{\Delta_{i_R}}\bigg)
\cr
& \times
 \prod_{i_L,j_L}\gamma_p\bigg(s_{i_L}s_{j_L} \big(1,\frac{1}{t_L^2}\big)_s x_{i_L j_L}^2\bigg)\prod_{i_R,j_R}\gamma_p\bigg(s_{i_R}s_{j_R}\big(1,\frac{1}{t_R^2}\big)_sx_{i_R j_R}^2  \bigg)\prod_{i_L,j_R}\gamma_p\bigg(s_{i_L}s_{j_R}x_{i_L j_R}^2  \bigg).
}
Using the $p$-adic Symanzik star integration formula \eno{SymanzikStar} to further simplify the pre-amplitude, we obtain
\eqn{}{
\tilde{\mathcal{A}}^{\text{exc}} &=
\zeta_p(1)^2
\zeta_p\big(\sum \Delta_{i_L}-\frac{n}{2}-c\big)\zeta_p\big(\sum \Delta_{i_R}-\frac{n}{2}+c\big)
\int[d\gamma] \prod_{i<j} \bigg[\frac{\zeta(2\gamma_{ij})}{|x_{ij}|_p^{2\gamma_{ij}}}\bigg]
\cr
&
\times \int_{\mathbb{Q}_{p}} \frac{dt_L}{|t_L|_p}
\int_{\mathbb{Q}_{p}}\frac{dt_R}{|t_R|_p}
|t_L|^{\frac{n}{2}-\sum\Delta_{i_L}-c} 
|t_R|^{\frac{n}{2}-\sum\Delta_{i_R}+c}
\left|1,\frac{1}{t_L}\right|_s^{-2\sum_{i_L<j_L}\gamma_{i_Lj_L}}\left|1,\frac{1}{t_R}\right|_s^{-2\sum_{i_R<j_R}\gamma_{i_Rj_R}}
}
where $[d\gamma]$ is defined in \eqref{GammaMeasure}. Further, in the following, we will often abbreviate sums like $\sum_{i_L<j_L}\gamma_{i_Lj_L}$ with $\sum\gamma_{i_Lj_L}$, so that such sums do not double-count terms. For the $t_L$ and $t_R$ integrals, one may note that
\eqn{tint}{
\int_{\mathbb{Q}_p}\frac{dt}{|t|_p}|t|_p^a\left|1,t\right|^b_s=\frac{\zeta_p(a)\zeta_p(-a-b)}{\zeta_p(1)\zeta_p(-b)}\,,
}
using which we conclude that
\eqn{exchPre1}{
\tilde{\mathcal{A}}^{\text{exc}} &=
\int[d\gamma]\; \prod_{i<j} \bigg[\frac{\zeta(2\gamma_{ij})}{|x_{ij}|_p^{2\gamma_{ij}}}\bigg]
\zeta_p\big(\sum \Delta_{i_L}-\frac{n}{2}-c\big)\zeta_p\big(\sum \Delta_{i_R}-\frac{n}{2}+c\big)
\cr
& \quad \times
\frac{\zeta_p(\sum\Delta_{i_L}-\frac{n}{2}+c)\zeta_p(2\sum\gamma_{i_Lj_L}-\sum\Delta_{i_L}+\frac{n}{2}-c)}{\zeta_p(2\sum\gamma_{i_Lj_L})} \cr 
&\quad \times
\frac{\zeta_p(\sum\Delta_{i_R}-\frac{n}{2}-c)\zeta_p(2\sum\gamma_{i_Rj_R}-\sum\Delta_{i_R}+\frac{n}{2}+c)}{\zeta_p(2\sum\gamma_{i_Rj_R})}\,.
}
Having worked out the pre-amplitude, all that remains in determining the Mellin exchange amplitude is to carry out the contour integral in \eqref{ExcContour}. Because of the delta functions in the integration measure $[d\gamma]$ given in \eno{GammaMeasure}, on the support of the integrand we have that 
\eqn{}
{
\sum\gamma_{i_Lj_L}-\sum\Delta_{i_L}=\sum\gamma_{i_Rj_R}-\sum\Delta_{i_R}\,.
}
The contour integral we need to compute over the complex cylinder can be evaluated using the identity
\eqn{contourId}
{
\frac{\log p}{2\pi i}\int_{-\frac{i\pi}{\log p}}^{\frac{i\pi}{\log p}}dc
&\,
\frac{\zeta_p(A+c)\zeta_p(A-c)\zeta_p(B+c)\zeta_p(B-c)\zeta_p(C+c)\zeta_p(C-c)\zeta_p(D+c)\zeta_p(D-c)}{\zeta_p(2c)\zeta_p(-2c)}
\cr
&=
2\,\frac{\zeta_p(A+B)\zeta_p(A+C)\zeta_p(A+D)\zeta_p(B+C)\zeta_p(B+D)\zeta_p(C+D)}{\zeta_p(A+B+C+D)}\,,
}
which, assuming $A,B,C,D>0$, can be straightforwardly verified, e.g.\ by closing the contour to the right and summing over the residues of the poles at $c$ equal to $A$, $B$, $C$, and $D$.
Using \eno{contourId}, we arrive at the result
\eqn{}
{
\mathcal{A}^{\text{exc}} &=
\zeta_p(\sum\Delta_i-n)\zeta_p(\Delta_A+\sum\Delta_{i_L}-n)\zeta_p(\Delta_A+\sum\Delta_{i_R}-n)
\cr
&\quad \times \int[d\gamma] \prod_{i<j} \bigg[\frac{\zeta_p(2\gamma_{ij})}{|x_{ij}|_p^{2\gamma_{ij}}}\bigg]\frac{\zeta_p(2\sum\gamma_{i_Lj_L}+\Delta_A-\sum\Delta_{i_L})}{\zeta_p(2\sum\gamma_{i_Lj_L}+\Delta_A+\sum\Delta_{i_R}- n)}\,,
}
from which we extract the Mellin amplitude,
\eqn{SingleFinal0}
{
\mathcal{M}^{\text{exc}} &=
\zeta_p(\sum\Delta_i-n)\zeta_p(\Delta_A+\sum\Delta_{i_L}-n)\zeta_p(\Delta_A+\sum\Delta_{i_R}-n)
\cr
&\quad \times \frac{\zeta_p(2\sum\gamma_{i_Lj_L}+\Delta_A-\sum\Delta_{i_L})}{\zeta_p(2\sum\gamma_{i_Lj_L}+\Delta_A+\sum\Delta_{i_R}- n)}\,.
}
It is instructive to write the Mellin amplitude in an alternate mathematically equivalent form,
\eqn{SingleFinal}
{
\mathcal{M}^{\text{exc}} &=
-\zeta_p(\Delta_A+\sum\Delta_{i_L}-n)\zeta_p(\Delta_A+\sum\Delta_{i_R}-n)
\cr
&\quad \times \left(\zeta_p(\sum \Delta_{i_L} -2\sum \gamma_{i_L j_L} - \Delta_A) - \zeta_p(\sum \Delta_i - n) \right) .
}
Unlike the contact Mellin amplitude, we see that the exchange diagram Mellin amplitude has explicit dependence on Mellin variables $\{\gamma_{i_Lj_L}\}$ via the Mandelstam variable (see \eno{MandelstamDef})
\eqn{sLexchDef}{
s_L \equiv s_{\{i_L\}} =  \sum \Delta_{i_L} -2\sum \gamma_{i_L j_L}\,.
}
We remind the reader that the sum in the first term in the final equality above is over all possible values that the index $i_L$ can take, i.e.\ all external legs to the left of the internal line, and the sum in the second term is over all such $i_L$ and $j_L$ with the condition $i_L < j_L$.
Finally, we note that we can readily extract the Mellin-Barnes integral representation of the exchange Mellin amplitude as quoted in \eno{padicExchExp} by comparing \eno{ExcContour} and \eno{exchPre1} with \eno{pMel}.

\subsection{Diagrams with two internal lines}
\label{doubleSection}

Next we consider a generic bulk diagram with two internal lines:
\eqn{}{
\begin{matrix}
\includegraphics[height=16ex]{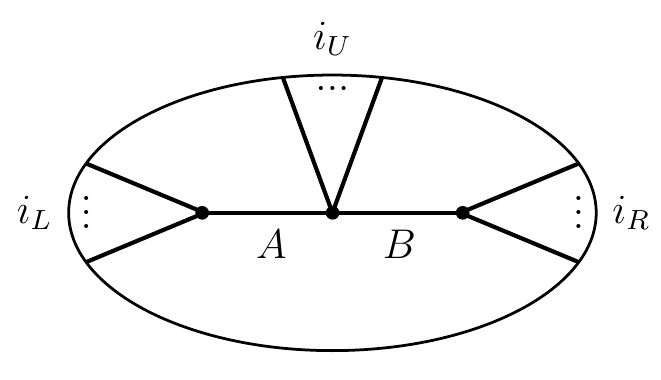}
 \end{matrix}\,.
}
Concretely, in terms of a product over propagators with three dummy bulk vertices summed over the entire Bruhat-Tits tree, the position space amplitude $\mathcal{A}^{2-\text{int}}$ is defined to be
\eqn{}
{
&
\sum_{(z^L_0,z^L),(z^U_0,z^U),(z^R_0,z^R)\in \mathcal{T}_{p^n}} \left(\prod_{i_L}K_{\Delta_{i_L}}(z^L_0,z^L;x_{i_L})\right)
G_{\Delta_A}(z^L,z^L_0;z^U,z^U_0)
\cr
&\hspace{36mm}
\times \left(\prod_{i_U} K_{\Delta_{i_U}}(z^C_0,z^C;x_{i_U})\right)
G_{\Delta_B}(z^U,z^U_0;z^R,z^R_0)
\left( \prod_{i_R} K_{\Delta_{i_R}}(z^R_0,z^R;x_{i_R})\right),
}
where $i_L$ runs over external legs on the left of the diagram, $i_R$ runs over external legs to the right, and $i_U$ runs over external legs incident to the centre vertex of the diagram.
Applying the split representation to, say, the $\Delta_A$ bulk-to-bulk propagator, the diagram decomposes into a contour integral over a contact diagram times an exchange diagram. 
Applying the results for contact and exchange amplitudes \eqref{preMellin} and \eqref{preMellinEx} from above to these components, we may re-write the position space amplitude as a contour integral of a certain ratio of local zeta functions times a pre-amplitude, that is,
\eqn{DoubleExcContour}{
\mathcal{A}^{2-\text{int}}=
&
\prod_{I\in \{A,B\}}
\bigg[\frac{1}{2}\frac{\log p}{2\pi i} 
\int_{-\frac{i\pi}{\log p}}^{\frac{i\pi}{\log p}}\, dc_I\,\frac{\zeta_p\big(\Delta_I-\frac{n}{2}+c_I\big)\zeta_p\big(\Delta_I-\frac{n}{2}-c_I\big)}{\zeta_p(2c_I)\zeta_p(-2c_I)}\bigg]
\tilde{\mathcal{A}}^{2-\text{int}}\,,
}
with the pre-amplitude $\tilde{\mathcal{A}}^{2-\text{int}}$ given by
\eqn{LongDouble}
{
&
\int_{\mathbb{Q}_{p^n}}dx_L
\sum_{a_L\in\{1,p\}}\zeta_p\big(\sum \Delta_{i_L}-\frac{n}{2}-c_A\big)
\prod_{i_L}\bigg(\frac{2\zeta_p(1)}{|2|_p}\int_{a_L\mathbb{Q}^2_p}\frac{ds_{i_L}}{|s_{i_L}|_p}|s_{i_L}|_p^{\Delta_{i_L}}\bigg)\prod_{i_L < j_L}\gamma_p\bigg(s_{i_L}s_{j_L}x_{i_Lj_L}^2  \bigg)
\cr
& \times 
\frac{2\zeta_p(1)}{|2|_p}\int_{a_L\mathbb{Q}^2_p} \frac{du}{|u_L|_p}\,|u_L|_p^{\frac{n}{2}-c_A} \prod_i \gamma_p\bigg(u_Ls_{i_L}(x_{i_L}-x_L)^2\bigg)
\cr
& \times 
\zeta_p\big(\sum \Delta_{i_C}+c_A-c_B\big)
\zeta_p\big(\sum \Delta_{i_R}-\frac{n}{2}+c_B\big)
\int_{\mathbb{Q}_p} \frac{dt_U}{|t_U|_p}\frac{dt_R}{|t_R|_p}|t_U|_p^{-\sum\Delta_{i_U}-c_A-c_B} 
|t_R|_p^{\frac{n}{2}-\sum\Delta_{i_R}+c_B}
\cr
& \times 
\zeta_p(1)^2\sum_{a_R\in\{1,p\}}
\prod_{i_U}\bigg(\frac{2\zeta_p(1)}{|2|}\int_{a_R\mathbb{Q}^2_p}\frac{ds_{i_U}}{|s_{i_U}|_p}|s_{i_U}|_p^{\Delta_{i_U}}\bigg)
\prod_{i_R}\bigg(\frac{2\zeta_p(1)}{|2|_p}\int_{a_R\mathbb{Q}^2_p}\frac{ds_{i_R}}{|s_{i_R}|_p}|s_{i_R}|_p^{\Delta_{i_R}}\bigg)
\cr
& \times 
\frac{2\zeta_p(1)}{|2|_p}
\int_{a_R\mathbb{Q}^2_p}\frac{du_U}{|u_U|_p} |u_U|_p^{\frac{n}{2}+c_B}
\prod_{i_U<j_U}\gamma_p\bigg(s_{i_U}s_{j_U} \big(1,\frac{1}{t_U^2}\big)_s x_{i_Uj_U}^2\bigg)\prod_{i_R<j_R}\gamma_p\bigg(s_{i_R}s_{j_R}\big(1,\frac{1}{t_R^2}\big)_sx_{i_Rj_R}^2  \bigg)
\cr
& \times 
\prod_{i_U,\, j_R}\gamma_p\bigg(s_{i_U}s_{j_R}x_{i_Uj_R}^2  \bigg)
\prod_{i_U}\gamma_p\bigg(u_Us_{i_U}(x_L-x_{i_U})^2\big(1,\frac{1}{t_U^2}\big)_s\bigg)\prod_{i_R}\gamma_p\bigg(u_Us_{i_R}(x_L-x_{i_R})^2\bigg)\,.
}
 In this form \eno{LongDouble} the symmetry with respect to the two internal propagators of the diagram is no longer apparent, but it will become manifest later. Changing to variables $S_{i_L}=\frac{s_{i_L}}{u_L}$, $S_{i_U}=\frac{s_{i_U}}{u_U(1,t_U^{-2})_s}$, and $S_{i_R}=\frac{s_{i_R}}{u_U}$,  the $x_L$ integral can be carried out just as in sections \ref{CONTACT} and \ref{singleSection} leading to a result that makes explicit reference to an index $m$ given by $|S_m|_p=\sup|S_i|_p$ (where $i$ now runs over all values that $i_L$, $i_U$, and $i_R$ take).  But just like in those sections, one can then do a change of variables from $S_i$ to new variables $s_i$, which eliminates explicit reference to the index $m$.  The pre-amplitude is then  expressed as 
\eqn{preMellinDouble}
{
\tilde{{\cal A}}^{2-\rm int} & = 
\zeta_p\big(\sum \Delta_{i_L}-\frac{n}{2}-c_A\big)
\zeta_p\big(\sum \Delta_{i_C}+c_A-c_B\big)
\zeta_p\big(\sum \Delta_{i_R}-\frac{n}{2}+c_B\big)
\cr
& \times \sum_{a\in\{1,p\}}
\prod_{i}\bigg(\frac{2\zeta_p(1)}{|2|_p}\int_{a\mathbb{Q}^2_p}\frac{ds_i}{|s_i|_p}|s_i|_p^{\Delta_i}\bigg) \zeta_p(1)^4\int_{\mathbb{Q}_p} \frac{du_L}{|u_L|_p}\frac{dt_U}{|t_U|_p}\frac{dt_R}{|t_R|_p}\frac{du_U}{|u_U|_p} \cr 
& \times 
|u_L|_p^{\sum\Delta_{i_L}-\frac{n}{2}+c_A}
|t_U|_p^{\sum\Delta_{i_U}+c_A+c_B} 
|t_R|_p^{\sum\Delta_{i_R}-\frac{n}{2}-c_B}
|u_U|_p^{\sum\Delta_{i_U}+\sum\Delta_{i_R}-\frac{n}{2}-c_A}
\cr
& \times 
\prod_{i_L< j_L}\gamma_p\bigg(s_{i_L}s_{j_L}\left(1,u_L\right)^2_sx_{i_Lj_L}^2  \bigg)
 \prod_{i_U<j_U}\gamma_p\bigg(s_{i_U}s_{j_U}\left(1,u_U,u_Ut_U,t_U^2\right)^2_sx_{i_Uj_U}^2 \bigg)
\cr
& \times 
 \prod_{i_L,i_U} \gamma_p\bigg(s_{i_L}s_{i_U}(1,t_U)^2_sx_{i_Li_U}^2\bigg)
 \prod_{i_L,j_R} \gamma_p\bigg(s_{i_L}s_{j_R}x_{i_Lj_R}^2\bigg)
 \prod_{i_U,j_R}\gamma_p\bigg(s_{i_U}s_{j_R}\left(1,u_U,t_U\right)^2_sx_{i_Uj_R}^2  \bigg)
  \cr 
& \times \prod_{i_R<j_R}\gamma_p\bigg(s_{i_R}s_{j_R}\left(1,u_U,u_Ut_R\right)^2_sx_{i_Rj_R}^2  \bigg)\,.
}
 Using the Symanzik star integration formula \eqref{SymanzikStar}, the pre-amplitude can now be written with the integrals over $s_i$ variables replaced by integrals over the Mellin variables $\gamma_{ij}$.
The remaining integrals over $u_L$, $t_U$, $t_R$, and $u_U$ still need to be worked out. After using \eno{SymanzikStar}, the $u_L$ integral factors out, and with a suitable change of variables, the $t_R$ integral can also be made to factor out. Both these integrals can then be immediately performed using \eqref{tint}, resulting in
\eqn{ApreampI}
{
\tilde{\mathcal{A}}^{2-\text{int}} &=
\zeta_p\big(\sum \Delta_{i_L}-\frac{n}{2}-c_A\big)
\zeta_p\big(\sum \Delta_{i_U}+c_A-c_B)\zeta(\sum \Delta_{i_R}-\frac{n}{2}+c_B\big)
\cr
& \times 
\int [d\gamma]\prod_{i, j}\left[\frac{\zeta_p(2\gamma_{ij})}{|x_{ij}|_p^{2\gamma_{ij}}}\right]
\frac{\zeta_p(\sum\Delta_{i_L}-\frac{n}{2}+c_A)\zeta_p(2\sum\gamma_{i_Lj_L}-\sum\Delta_{i_L}+\frac{n}{2}-c_A)}{\zeta_p(2\gamma_{i_Lj_L})}
\cr
& \times 
\frac{\zeta_p(\sum\Delta_{i_R}-\frac{n}{2}-c_B)\zeta_p(2\sum\gamma_{i_Rj_R}-\sum\Delta_{i_R}+\frac{n}{2}+c_B)}{\zeta_p(2\gamma_{i_Rj_R})} 
\,
\mathcal{I}(s_L,s_R,\Delta_{i_U},c_A,c_B)\,,
}
where we have lumped together the remaining $t_U$ and $u_U$ integrals into $\mathcal{I}(s_L,s_R,\Delta_{i_U},c_A,c_B)$, defined to be
\eqn{Idef}
{
\mathcal{I}(s_L,s_R,\Delta_{i_U},c_A,c_B)&\equiv
\zeta_p(1)^2
\int_{\mathbb{Q}_p} \frac{dt_U}{|t_U|_p}|t_U|_p^{\sum\Delta_{i_U}+c_A+c_B} 
\int_{\mathbb{Q}_p}\frac{du_U}{|u_U|_p} |u_U|_p^{\sum\Delta_{i_U}+c_B-c_A}
\cr
& \hspace{4.5mm}\times 
|1,t_U|_s^{s_R-s_L-\sum \Delta_{i_U}}
|1,u_U,t_U|_s^{s_L-s_R-\sum \Delta_{i_U}}
|1,u_U|_s^{s_R-\frac{n}{2}-c_B}\,,
}
and we have identified the Mandelstam-like variables
\eqn{MandelstamLR}
{
s_L \equiv \sum\Delta_{i_L} - 2\sum\gamma_{i_Lj_L} \qquad 
 s_R \equiv \sum\Delta_{i_R} - 2\sum\gamma_{i_Rj_R}\,,
}
where like before, it is understood that in the sum over Mellin variables $\gamma_{i_L j_L}$ ($\gamma_{i_R j_R}$) the sum is restricted to $i_L < j_L$ ($i_R < j_R$). The Mandelstam variables satisfy
\eqn{}{
s_L &= s_R + \sum \Delta_{i_U} -2\sum \gamma_{i_U j_U} - 2\sum \gamma_{i_U i_R} \cr 
s_R &= s_L + \sum \Delta_{i_U} -2\sum \gamma_{i_U j_U} - 2\sum \gamma_{i_U i_L}\,, 
}
where the sum over the Mellin variables $\gamma_{i_U j_L}$ and $\gamma_{i_U j_R}$ is unrestricted in the indices.
This integral can be performed by, for example, partitioning the integration domain into regions where $t_U$, $u_U$ or $1$ have the largest $p$-adic norm, thus simplifying the integrand. We find 
\eqn{Iexpression}
{
\mathcal{I}(s_L,s_R,\Delta_{i_U},c_A,c_B)  &=
\zeta_p(s_L-\frac{n}{2}-c_A)\zeta_p(s_R-\frac{n}{2}+c_B) 
\cr
&
-\zeta_p\big(\sum\Delta_{i_U}+c_A+c_B\big)\bigg[
\zeta_p(s_L-\frac{n}{2}-c_A)-\zeta_p\big(\sum\Delta_{i_U} + c_B-c_A\big)
\bigg]
\cr
&
-\zeta_p\big(\sum\Delta_{i_U}-c_A-c_B\big)\bigg[
\zeta_p(s_R-\frac{n}{2}+c_B)-\zeta_p\big(\sum\Delta_{i_U} + c_B-c_A\big)
\bigg]
\cr
&
-\zeta_p\big(\sum\Delta_{i_U} + c_B-c_A\big)\,.
}

With the pre-amplitude in hand, we are ready to carry out the two contour integrals in \eqref{DoubleExcContour} to obtain the full Mellin amplitude. 
One way to do this is to close both contours to the right, and sum over the residues in the $c_A$ plane, which occur at
\eqn{}{
c_A =\left\{ \Delta_A-\frac{n}{2}\,, \sum\Delta_{i_L}-\frac{n}{2}\,, 2\sum\gamma_{i_Lj_L}-\sum\Delta_{i_L}+\frac{n}{2}\,, \sum\Delta_{i_U}+c_B, \sum\Delta_{i_U}-c_B \right\},
}
and then sum over the residues in the $c_B$ plane, occurring at
\eqn{}{
c_B &= \left\{ \Delta_B-\frac{n}{2}\,, \sum\Delta_{i_L}+\sum\Delta_{i_U}-\frac{n}{2}\,, \sum\Delta_{i_U}+\Delta_B-\frac{n}{2}\,, \sum\Delta_{i_R}-\frac{n}{2}\,, \right. \cr 
& \qquad \left. 2\sum\gamma_{i_Rj_R}-\sum\Delta_{i_R}+\frac{n}{2} \right\}.
}
We omit the details of this step, which leads to  the final expression for the diagram with two internal lines. From this we easily extract the closed-form expression for the Mellin amplitude,
\eqn{DoubleFinalA}{
\mathcal{M}^{2-\text{int}} &=
\zeta_p\big(\sum\Delta_{i_L}+\Delta_A-n\big)
\zeta_p\big(\sum\Delta_{i_U}+\Delta_A+\Delta_B-n\big)
\zeta_p\big(\sum\Delta_{i_R}+\Delta_B-n\big)
\cr
& \times
\bigg[
\zeta_p(s_L-\Delta_A)\zeta_p(s_R-\Delta_B) -\zeta_p\big(\sum\Delta_i-n\big) \cr 
& \quad -\zeta_p\big(\sum \Delta_{i_R} +\sum\Delta_{i_U}+\Delta_A-n\big)
\bigg(\zeta_p(s_L-\Delta_A)-\zeta_p\big(\sum\Delta_i-n\big)\bigg)
\cr
& 
\quad -
\zeta_p\big(\sum\Delta_{i_U}+\sum \Delta_{i_L} +\Delta_B-n\big)
\bigg(\zeta_p(s_R-\Delta_B)-\zeta_p\big(\sum\Delta_i-n\big)\bigg)
\bigg]\,.
}
The Mellin-Barnes integral representation of this amplitude  may be easily extracted from \eno{DoubleExcContour}, \eno{ApreampI} and \eno{Iexpression}.

\subsection{Diagrams with three internal lines}
\label{tripleSection}

Finally, we now provide a first principles derivation of the Mellin amplitudes of the bulk diagrams  with three internal lines. 
The Mellin amplitudes of these diagrams can be computed using essentially the same methods by which the exchange diagram and the diagram with two internal lines were derived above, although the intermediate steps are more cumbersome.  One new feature, though, that appears at three internal lines is the existence of two different diagrammatic topologies: the three internal lines can be arranged in series or meet at a centre vertex.

Using the split representation of the bulk-to-bulk propagator on an internal leg, diagrams with three internal lines can be split into the product of a contact diagram and a diagram with two internal lines or two diagrams each with one internal line, and these two diagrams are connected via a boundary integral. Applying equation \eqref{DoubleExcContour} to the component with two internal legs then leads to the representation
\eqn{TripleExcContour}{
\mathcal{A}^{3-\text{int}}=
&
\prod_{I=A,B,C}
\bigg[\frac{1}{2}\frac{\log p}{2\pi i} 
\int_{-\frac{i\pi}{\log p}}^{\frac{i\pi}{\log p}}\, dc_I\,\frac{\zeta_p\big(\Delta_I-\frac{n}{2}+c_I\big)\zeta_p\big(\Delta_I-\frac{n}{2}-c_I\big)}{\zeta_p(2c_I)\zeta_p(-2c_I)}\bigg]
\tilde{\mathcal{A}}^{3-\text{int}},
}
where the pre-amplitude $\tilde{\mathcal{A}}^{3-\text{int}}$ can be found by invoking equations \eqref{preMellin} and \eqref{preMellinDouble}. As for the exchange diagrams from section \ref{singleSection}, one can, by performing a series of suitable change of variables, carry out the boundary integral that connects the contact diagram and  two-internal-line diagram components of the Mellin amplitude $\mathcal{M}^{3-\text{int}}$, and then use the  Symanzik star integration formula \eno{SymanzikStar} to write the pre-amplitude as a Mellin integral. Thereafter, one will need to carry out six integrals over auxiliary variables, similar to the $u_L$, $t_U$, $t_R$, and $u_U$ integrals from section \ref{doubleSection},  to obtain the final result for $\mathcal{M}^{3-\text{int}}$. 

We demonstrate this procedure explicitly for diagrams with three internal lines, starting with the diagram where the internal lines arrange in a series configuration.

\subsubsection*{Diagram with three internal lines in a series.}

The arbitrary-point diagram with three internal lines arranged in a series is represented diagrammatically as
\eqn{TripDiagram}{
\begin{matrix}
\includegraphics[height=16ex]{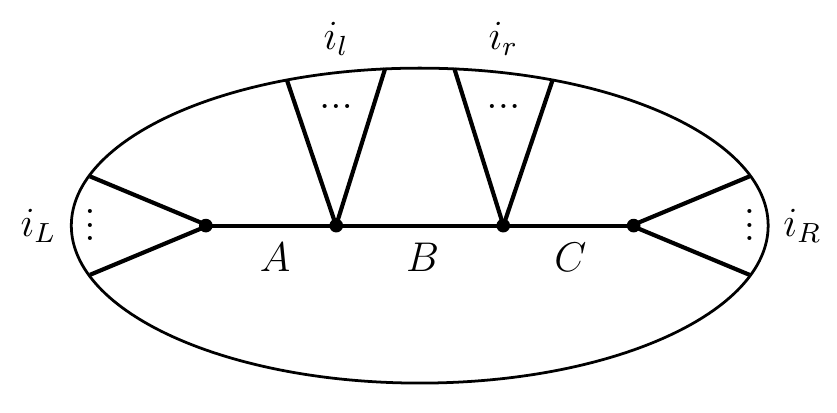}
 \end{matrix}\,.
}
Written explicitly in terms of bulk-to-bulk and bulk-to-boundary propagators, this diagram is given by
\eqn{}
{
\mathcal{A}^{3-\text{int, line}}
&= 
\sum_{\substack{(z^L_0,z^L),(z^l_0,z^l),\\(z^R_0,z^R),(z^r_0,z^r)\in \mathcal{T}_{p^n}}} 
\left( \prod_{i_L} K_{\Delta_{i_L}}(z^L_0,z^L;x_{i_L})\right)
G_{\Delta_A}(z^L,z^L_0;z^l,z_0^l)
\cr
&  
\hspace{29mm}
\times \left(\prod_{i_l} K_{\Delta_{i_l}}(z^l_0,z^l;x_{i_l})\right)
G_{\Delta_B}(z^l,z^l_0;z^r,z_0^r)
\cr
& 
\hspace{29mm}
\times \left(\prod_{i_r} K_{\Delta_{i_r}}(z^r_0,z^r;x_{i_r})\right)
G_{\Delta_C}(z^r,z^r_0;z^R,z_0^R)
\cr
& 
\hspace{29mm}
\times \left(\prod_{i_R} K_{\Delta_{i_R}}(z^R_0,z^R;x_{i_R})\right),
}
where the summation symbol in front denotes the four bulk integrations (more precisely, tree summations) over the four bulk vertices, and the indices $i_L, i_l, i_r, i_R$ run over different external legs as depicted in \eno{TripDiagram}.

The pre-amplitude for this diagram is given by
\eqn{TripleLinePreAmp}
{
\tilde{\mathcal{A}}^{3-\text{int, line}} &=
\int [d\gamma]\prod_{i<j}\bigg[\frac{\zeta(2\gamma_{ij})}{|x_{ij}|_p^{2\gamma_{ij}}}\bigg]
\mathcal{I}(s_L,s_c,\Delta_l,c_A,c_B)
\,
\mathcal{I}(s_c,s_R,\Delta_r,c_B,c_C)
\cr
& \times 
\zeta_p\big(\Delta_{L}-\frac{n}{2}-c_A\big)
\zeta_p(\Delta_{l}+c_A-c_B)
\zeta_p(\Delta_{r}+c_B-c_C)
\zeta_p\big(\Delta_{R}-\frac{n}{2}+c_C\big)
\cr
& \times
\bigg(\zeta_p\big(\Delta_{L}-\frac{n}{2}+c_A\big)-\zeta_p\big(s_{L}-\frac{n}{2}+c_A\big)\bigg)
\bigg(\zeta_p\big(\Delta_{R}-\frac{n}{2}-c_C\big)-\zeta_p\big(s_{R}-\frac{n}{2}-c_C\big)\bigg),
}
where the function $\mathcal{I}$ is given in equation \eqref{Iexpression}, the Mandelstam invariants $s_L$ and $s_R$ of the left and right legs are given in \eqref{MandelstamLR}, while that of the center internal leg is given by
\eqn{MandelstamSeries}{
 s_c &= 
 \Delta_{L}+\Delta_{l}
 - 2\sum_{i_L < j_L}\gamma_{i_Lj_L}
 - 2\sum_{i_L , j_l}\gamma_{i_Lj_l}
 - 2\sum_{i_l < j_l}\gamma_{i_lj_l}
\cr
&= 
  \Delta_{r}+\Delta_{R}
 - 2\sum_{i_r < j_r}\gamma_{i_rj_r}
 - 2\sum_{i_r , j_R}\gamma_{i_rj_R}
 - 2\sum_{i_R < j_R}\gamma_{i_Rj_R}\,.
}
In \eno{TripleLinePreAmp}-\eno{MandelstamSeries} we have introduced a shortened notation,
\eqn{}
{
&
\Delta_L \equiv \sum_{i_L}\Delta_{i_L}
\hspace{10mm}
\Delta_l \equiv \sum_{i_l}\Delta_{i_l}
\hspace{10mm}
\Delta_r \equiv \sum_{i_r}\Delta_{i_r}
\hspace{10mm}
\Delta_R \equiv \sum_{i_R}\Delta_{i_R}\,.
}
One can carry out the three contour integrals over the pre-amplitude by closing all contours to the right and summing over the residues at
\eqn{}{
c_A = \left\{ \Delta_A-\frac{n}{2}; \Delta_L-\frac{n}{2}; \frac{n}{2}-s_L; \Delta_l+c_B; \Delta_l-c_B\right\},
}
and then summing over the residues at 
\eqn{}{
c_C = \left\{ \Delta_C-\frac{n}{2}; \Delta_R-\frac{n}{2}; \frac{n}{2}-s_R; \Delta_r+c_B; \Delta_r-c_B \right\},
}
followed by summing over the residues at 
\eqn{}{
c_B = \left\{ \Delta_B-\frac{n}{2}; \Delta_{A}+\Delta_{l}-\frac{n}{2}; \Delta_l+\Delta_L-\frac{n}{2}; \frac{n}{2}-s_c; \Delta_C+\Delta_r-\frac{n}{2}; \Delta_r+\Delta_R-\frac{n}{2} \right\}.
}
This leads to the final result for the Mellin amplitude, 
\eqn{TripleLineAmp}{
\mathcal{M}^{3-\text{int, series}}=
-\zeta_p&\big(\Delta_{AL,}-n\big)
\zeta_p\big(\Delta_{ABl,}-n\big)
\zeta_p\big(\Delta_{BCr,}-n\big)
\zeta_p\big(\Delta_{CR,}-n\big)
\cr
\Bigg\{ 
&
\zeta_p(s_L-\Delta_A)\zeta_p(s_c-\Delta_B)\zeta_p(s_R-\Delta_C) 
-\zeta_p(\textstyle \sum\Delta_{i}-n)
\cr 
&
-\zeta_p(\Delta_{BLl,}-n) \bigg[ \zeta_p(s_c-\Delta_B)\zeta_p(s_R-\Delta_C)-\zeta_p(\textstyle \sum\Delta_{i}-n)  
\cr 
&
\hspace{30mm}
- \zeta_p(\Delta_{CLlr,}-n) \Big( \zeta_p(s_R-\Delta_C) - \zeta_p(\textstyle \sum\Delta_{i}-n) \Big)
\cr 
&
\hspace{31mm}
- \zeta_p(\Delta_{BrR,}-n) \Big( \zeta_p(s_c-\Delta_B) - \zeta_p(\textstyle \sum\Delta_{i}-n)  \Big) \bigg] 
\cr 
&
-\zeta_p(\Delta_{BrR,}-n) 
\bigg[ \zeta_p(s_L-\Delta_A)\zeta_p(s_c-\Delta_B) -\zeta_p(\textstyle \sum\Delta_{i}-n)
\cr
&
\hspace{30mm}- \zeta_p(\Delta_{AlrR,}-n) 
\Big( \zeta_p(s_L-\Delta_A)- \zeta_p(\textstyle \sum\Delta_{i}-n) \Big) 
\cr 
&
\hspace{31mm}  - \zeta_p(\Delta_{BLl,}-n) 
\Big( \zeta_p(s_c-\Delta_B) - \zeta_p(\textstyle \sum\Delta_{i}-n) \Big) \bigg] 
\cr 
&
-\zeta_p(\Delta_{AClr,}-n) 
\bigg[ \zeta_p(s_L-\Delta_A)\zeta_p(s_R-\Delta_C)  
-\zeta_p(\textstyle \sum\Delta_{i}-n)
\cr
&
\hspace{30mm}
- \zeta_p(\Delta_{CLlr,}-n) \Big( \zeta_p(s_R-\Delta_C)  - \zeta_p(\textstyle \sum\Delta_{i}-n) \Big) 
\cr 
&
\hspace{30mm}
 - \zeta_p(\Delta_{AlrR,}-n) \Big( \zeta_p(s_L-\Delta_A) - \zeta_p(\textstyle \sum\Delta_{i}-n)  \Big) 
 \bigg] 
\cr 
&
- \zeta_p(\Delta_{CLlr,}-n) 
\Big[ \zeta_p(s_R-\Delta_C)-\zeta_p(\textstyle \sum\Delta_{i}-n) \Big] 
\cr 
&
- \zeta_p(\Delta_{AlrR,}-n) 
\Big[ \zeta_p(s_L-\Delta_A)-\zeta_p(\textstyle \sum\Delta_{i}-n) \Big] 
\cr
&
- \zeta_p(\Delta_{BLl,}-n) \zeta_p(\Delta_{BrR,}-n) 
\Big[ \zeta_p(s_c-\Delta_B)-\zeta_p(\textstyle \sum\Delta_{i}-n) \Big] \Bigg\}\,. 
}

\subsubsection*{Star diagram with three internal lines.}

The star diagram, which is the other type of diagram with three internal lines, can be depicted diagrammatically as
\eqn{}{
\begin{matrix}
\includegraphics[height=24ex]{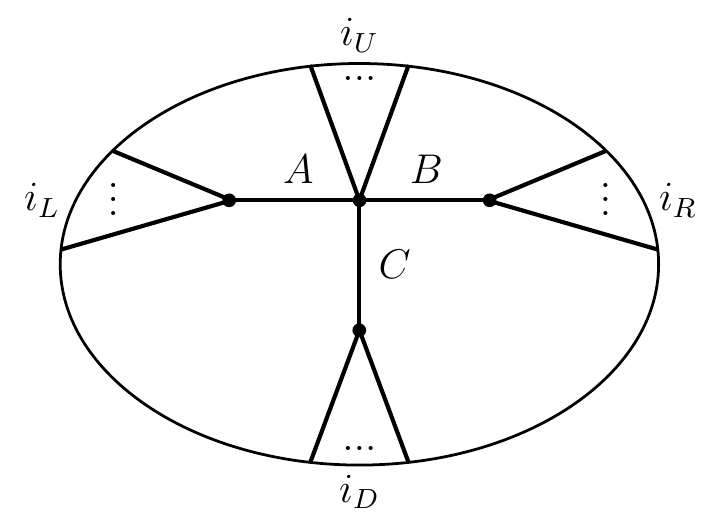}
 \end{matrix}\,.
}
Explicitly, this diagram corresponds to the position space amplitude,
\eqn{}
{
\mathcal{A}^{3-\text{int, star}}
 &=
\sum_{\substack{(z^U_0,z^U),(z^L_0,z^L),\\(z^R_0,z^R),(z^D_0,z^D)\in \mathcal{T}_{p^n}}} 
\left( \prod_{i_L} K_{\Delta_{i_L}}(z^L_0,z^L;x_{i_L})\right)
{G}_{\Delta_A}(z^L,z^L_0;z^U,z_0^U)
\cr
&
\hspace{29mm}
\times \left(\prod_{i_R} K_{\Delta_{i_R}}(z^R_0,z^R;x_{i_R})\right)
{G}_{\Delta_B}(z^R,z^R_0;z^U,z_0^U)
\cr
&
\hspace{29mm}
\times \left(\prod_{i_D} K_{\Delta_{i_D}}(z^C_0,z^C;x_{i_D})\right)
{G}_{\Delta_C}(z^D,z^D_0;z^U,z_0^U)
\cr
&
\hspace{29mm}
\times \left(\prod_{i_U} K_{\Delta_{i_U}}(z^U_0,z^U;x_{i_U})\right).
}
We introduce one more shorthand and a Mandelstam invariant,
\eqn{}{
\Delta_D \equiv \sum_{i_D}\Delta_{i_D}
\hspace{20mm}
 s_D &= 
 \Delta_{D}
 - 2\sum_{i_D < j_D}\gamma_{i_Dj_D}\,.
}
In terms of these, the pre-amplitude is given by
\eqn{StarPreAmp}{
\tilde{\mathcal{A}}^{3-\text{int, star}} &=
\zeta_p\big(\Delta_L-\frac{n}{2}-c_A\big)
\zeta_p\big(\Delta_R-\frac{n}{2}+c_B\big)
\zeta_p\big(\Delta_D-\frac{n}{2}-c_C\big)
\zeta_p\big(\Delta_U+\frac{n}{2}+c_C+c_A-c_B\big)
\cr
& \times 
\int [d\gamma] \prod_{i<j}\bigg[\frac{\zeta(2\gamma_{ij})}{|x_{ij}|_p^{2\gamma_{ij}}}\bigg]
\bigg(\zeta_p\big(\Delta_L-\frac{n}{2}+c_A\big)-\zeta_p\big(s_L-\frac{n}{2}+c_A\big)\bigg)
\,\mathcal{J}
\cr
& \times 
\bigg(\zeta_p\big(\Delta_D-\frac{n}{2}+c_C\big)-\zeta_p\big(s_D-\frac{n}{2}+c_C\big)\bigg)
\bigg(\zeta_p\big(\Delta_R-\frac{n}{2}-c_B\big)-\zeta_p\big(s_R-\frac{n}{2}-c_B\big)\bigg),
}
where $s_L, s_R$ are as given in \eno{MandelstamLR}.
The symbol $\mathcal{J}$ is a shorthand for an integral over the three auxiliary variables associated with the center vertex of the diagram. This integral is a more complicated version of the integral \eqref{Idef} and naturally appears if one attempts to compute the star diagram by the method described above. The integral is given by
\eqn{Jdef}
{
\mathcal{J} &\equiv \zeta_p(1)^3 \int_{\mathbb{Q}_p}\frac{dt}{|t|_p}\frac{du}{|u|_p}\frac{dm}{|m|_p}
|t|_p^{e_1}|u|_p^{e_2}|m|_p^{e_3}|1,t|_s^{e_4} |1,u|_s^{e_5}|1,m|_s^{e_6} |1,m,mu|_s^{e_7}|1,u,m,mu,mut|_s^{e_8}\,,
}
where the exponents assume the following values:
\eqn{}
{
& e_1= \Delta_U+\frac{n}{2}+c_A+c_B+c_C\,, \cr
& e_2= \Delta_U+\frac{n}{2}+c_B+c_C-c_A\,, \cr
& e_3= \Delta_U+\frac{n}{2}+c_c-c_a-c_b\,, \cr
& e_4= -\frac{n}{2}-\Delta_U+c_B-c_A-c_C\,, \cr
& e_5= -s_R-\Delta_U-c_A-c_C\,, \cr
& e_6= s_D-s_L-s_R-\Delta_U\,, \cr
& e_7= s_L-\frac{n}{2}+c_A\,, \cr
& e_8= s_R-\frac{n}{2}-c_B\,.
}
By carefully partitioning the domain of the integral \eqref{Jdef} according to which of $u$, $t$, $m$, or 1 has the biggest $p$-adic norm, one can explicitly compute $\mathcal{J}$ to find that
\eqn{}{
\mathcal{J}=
\Bigg\{ 
&
-\zeta_p(s_L-\frac{n}{2}-c_A)\zeta_p(s_R-\frac{n}{2}+c_B)\zeta_p(s_D-\frac{n}{2}-c_C) 
+\zeta_p(\Delta_{U}+\frac{n}{2}+c_{B,AC})
\cr 
&
+\zeta_p(\Delta_{U}+\frac{n}{2}+c_{A,BC}) 
\bigg[ 
\zeta_p(s_L-\frac{n}{2}-c_A)
\zeta_p(s_R-\frac{n}{2}+c_B)-\zeta_p(\Delta_{U}+\frac{n}{2}+c_{B,AC})
\cr 
&
\hspace{42mm}
- \zeta_p(\Delta_{U}+\frac{n}{2}+c_{AB,C}) \Big( \zeta_p(s_L-\frac{n}{2}-c_A) - \zeta_p(\Delta_{U}+\frac{n}{2}+c_{B,AC}) \Big)
\cr 
&
\hspace{42mm}
- \zeta_p(\Delta_{U}+\frac{n}{2}-c_{ABC,}) \Big( \zeta_p(s_R-\frac{n}{2}+c_B) - \zeta_p(\Delta_{U}+\frac{n}{2}+c_{B,AC}) \bigg)\bigg] 
\cr 
&
+\zeta_p(\Delta_{U}+\frac{n}{2}+c_{C,AB}) 
\bigg[ \zeta_p(s_R-\frac{n}{2}+c_B)\zeta_p(s_D-\frac{n}{2}-c_C) -\zeta_p(\Delta_{U}+\frac{n}{2}+c_{B,AC})
\cr
&
\hspace{42mm} - \zeta_p(\Delta_{U}+\frac{n}{2}-c_{ABC,}) 
\Big( \zeta_p(s_R-\frac{n}{2}+c_B) - \zeta_p(\Delta_{U}+\frac{n}{2}+c_{B,AC}) \Big) 
\cr 
&
\hspace{42mm} - \zeta_p(\Delta_{U}+\frac{n}{2}+c_{BC,A}) 
\Big( \zeta_p(s_D-\frac{n}{2}-c_C) - \zeta_p(\Delta_{U}+\frac{n}{2}+c_{B,AC}) \Big) \bigg] 
\cr 
&
+\zeta_p(\Delta_{U}+\frac{n}{2}+c_{ABC}) 
\bigg[ \zeta_p(s_L-\frac{n}{2}-c_A)\zeta_p(s_D-\frac{n}{2}-c_C)   
-\zeta_p(\Delta_{U}+\frac{n}{2}+c_{B,AC})
\cr
&
\hspace{42mm}
- \zeta_p(\Delta_{U}+\frac{n}{2}+c_{AB,C}) \Big( \zeta_p(s_L-\frac{n}{2}-c_A) - \zeta_p(\Delta_{U}+\frac{n}{2}+c_{B,AC}) \Big)
\cr 
&
\hspace{42mm}
- \zeta_p(\Delta_{U}+\frac{n}{2}+c_{BC,A}) \Big( \zeta_p(s_D-\frac{n}{2}-c_C) - \zeta_p(\Delta_{U}+\frac{n}{2}+c_{B,AC}) \Big) 
 \bigg] 
\cr 
&
+ \zeta_p(\Delta_{U}+\frac{n}{2}+c_{AB,C})
\Big[\zeta_p(s_L-\frac{n}{2}-c_A)-\zeta_p(\Delta_{U}+\frac{n}{2}+c_{B,AC}) \Big] 
\cr 
&
+ \zeta_p(\Delta_{U}+\frac{n}{2}-c_{ABC,}) 
\Big[\zeta_p(s_R-\frac{n}{2}+c_B)-\zeta_p(\Delta_{U}+\frac{n}{2}+c_{B,AC}) \Big] 
\cr
&
+ \zeta_p(\Delta_{U}+\frac{n}{2}+c_{BC,A}) 
\Big[\zeta_p(s_D-\frac{n}{2}-c_C)-\zeta_p(\Delta_{U}+\frac{n}{2}+c_{B,AC}) \Big] \Bigg\}\,,
}
where we used the short-hand
\eqn{cijkComma}{
c_{i_1 \ldots i_k,i_{k+1} \ldots i_\ell} \equiv \sum_{j=1}^k c_{i_j} - \sum_{j=k+1}^\ell c_{i_j}\,.
}
With the pre-amplitude $\tilde{\mathcal{A}}^{3-\text{int, star}}$ in hand, we are in a position to evaluate the three-fold contour integral in \eno{TripleExcContour}. The contour integral can, if one chooses to close the contour on the right, be computed by summing over the residues in the $c_C$ plane, at 
\eqn{cCRes}{
c_C = \bigg\{ &\Delta_C-\frac{n}{2}\,;\, 
\Delta_D-\frac{n}{2}\,;\, 
\frac{n}{2}-s_D\,;\, 
\Delta_U+\frac{n}{2}+c_{AB,}\,;\, 
\Delta_U+\frac{n}{2}+c_{A,B}\,;\,
\cr
&
\Delta_U+\frac{n}{2}+c_{B,A}\,;\,
\Delta_U+\frac{n}{2}-c_{AB,} \bigg\} ,
}
followed by summing over the residues in the $c_A$ plane, at 
\eqn{cARes}{
c_A = 
\left\{ \Delta_A-\frac{n}{2}\,;\,
\Delta_L-\frac{n}{2}\,;\,
\frac{n}{2}-s_L\,;\,
\Delta_{CU,}+c_B\,;\,
\Delta_{CU,}-c_B\,;\,
\Delta_{UD,}+c_B\,;\, 
\Delta_{UD,}-c_B\right\},
}
and finally summing over the residues in $c_B$ plane, at
\eqn{cBRes}{
c_B = \left\{ \Delta_B-\frac{n}{2}\,;\,
\Delta_R-\frac{n}{2}\,;\,
\frac{n}{2}-s_R\,;\,
\Delta_{ACU,}-\frac{n}{2}\,;\,
\Delta_{CLU,}-\frac{n}{2}\,;\,
\Delta_{AUD,}-\frac{n}{2}\,;\,
\Delta_{LUD,}-\frac{n}{2} \right\}.
}

When the dust settles, the Mellin amplitude of the star diagram is extracted to be
\eqn{StarAmp}{
\mathcal{M}^{3-\text{int, star}}=
\zeta_p&\big(\Delta_{AL,}-n\big)
\zeta_p\big(\Delta_{ABCU,}-n\big)
\zeta_p\big(\Delta_{CD,}-n\big)
\zeta_p\big(\Delta_{BR,}-n\big)
\cr
\Bigg\{ 
&
-\zeta_p(s_L-\Delta_A)\zeta_p(s_R-\Delta_B)\zeta_p(s_D-\Delta_C) 
+\zeta_p(\textstyle \sum\Delta_{i}-n)
\cr 
&
+\zeta_p(\Delta_{ABUD,}-n) \bigg[ \zeta_p(s_L-\Delta_A)\zeta_p(s_R-\Delta_B)-\zeta_p(\textstyle \sum\Delta_{i}-n)  
\cr 
&
\hspace{30mm}
- \zeta_p(\Delta_{AUDR,}-n) \Big( \zeta_p(s_L-\Delta_A) - \zeta_p(\textstyle \sum\Delta_{i}-n) \Big)
\cr 
&
\hspace{31mm}
- \zeta_p(\Delta_{BLUD,}-n) \Big( \zeta_p(s_R-\Delta_B) - \zeta_p(\textstyle \sum\Delta_{i}-n)\Big) \bigg] 
\cr 
&
+\zeta_p(\Delta_{BCUL,}-n) 
\bigg[ \zeta_p(s_R-\Delta_B)\zeta_p(s_D-\Delta_C) -\zeta_p(\textstyle \sum\Delta_{i}-n)
\cr
&
\hspace{30mm}- \zeta_p(\Delta_{BLUD,}-n) 
\Big( \zeta_p(s_R-\Delta_B)-\zeta_p(\textstyle \sum\Delta_{i}-n) \Big) 
\cr 
&
\hspace{31mm}  - \zeta_p(\Delta_{CLUR,}-n) 
\Big( \zeta_p(s_D-\Delta_C) - \zeta_p(\textstyle \sum\Delta_{i}-n)\Big) \bigg] 
\cr 
&
+\zeta_p(\Delta_{ACUR,}-n) 
\bigg[ \zeta_p(s_L-\Delta_A)\zeta_p(s_D-\Delta_C)  
-\zeta_p(\textstyle \sum\Delta_{i}-n)
\cr
&
\hspace{30mm}
- \zeta_p(\Delta_{AUDR,}-n) \Big( \zeta_p(s_L-\Delta_A)  - \zeta_p(\textstyle \sum\Delta_{i}-n)\Big)
\cr 
&
\hspace{30mm}
 - \zeta_p(\Delta_{CLUR,}-n) \Big( \zeta_p(s_D-\Delta_C) - \zeta_p(\textstyle \sum\Delta_{i}-n)
\Big) \bigg] 
\cr 
&
+ \zeta_p(\Delta_{AUDR,}-n) 
\Big[ \zeta_p(s_L-\Delta_A)-\zeta_p(\textstyle \sum\Delta_{i}-n) \Big] 
\cr 
&
+ \zeta_p(\Delta_{BLUD,}-n) 
\Big[ \zeta_p(s_R-\Delta_B)-\zeta_p(\textstyle \sum\Delta_{i}-n) \Big] 
\cr
&
+ \zeta_p(\Delta_{CLUR,}-n) 
\Big[ \zeta_p(s_D-\Delta_C)-\zeta_p(\textstyle \sum\Delta_{i}-n) \Big] \Bigg\}. 
}

\vspace{1em}
The pre-amplitudes of the previous two diagrams are easily read off of the intermediate steps of the derivation. Further, as can be seen, the amplitudes develop poles precisely when the Mandelstam-like variables equal  the dimension of the single-trace operators exchanged along the internal lines. While it is certainly possible to evaluate the $p$-adic Mellin amplitudes of tree-level bulk-diagrams with more than three internal lines using the techniques described in this section (and obtain closed-form expressions), we believe no fundamentally new tricks or techniques are required to extend the presentation of this section. 
One may wonder if there are other fundamentally different but more efficient techniques to reconstruct such Mellin amplitudes, such as perhaps recursion relations similar to the ones known for real Mellin amplitudes. 
The answer to this question turns out to be in the affirmative~\cite{Jepsen:2018toappear}.

\section{Outlook}
\label{DISCUSSION}

We have seen in this paper that Mellin space, which has proven to be a useful tool in the computation of correlators in conventional AdS/CFT, can also be defined in the context of $p$-adic AdS/CFT, where it proffers the same benefits compared with position space. For instance, arbitrary-point tree-level bulk diagrams can be evaluated relatively straightforwardly, are expressible in a compact form as meromorphic functions of Mellin variables, 
with poles corresponding to the exchange of solely single-trace operators.
We have also seen that the expressions for $p$-adic Mellin amplitudes exhibit a close resemblance to their real counterparts, sharing almost identical functional forms in the Mellin-Barnes contour integral representation, reflective of the fact that the intermediate steps of the computations closely parallel each other. 
Indeed, we have established the $p$-adic analogs of the split representation of the bulk-to-bulk propagator and the Symanzik star-integration formula, which are both used in the evaluation of bulk diagrams. 
One conspicuous difference, though, is that it is not necessary to pass to an embedding space formalism, due to the simple forms the bulk-to-bulk and bulk-to-boundary propagators already assume in $p$-adic AdS/CFT~\cite{Gubser:2016guj}.
Nevertheless, it would be interesting to undertake a closer analysis of a $p$-adic analog of the embedding space formalism -- which over the reals  owes its existence to the Euclidean $n$-dimensional conformal algebra $SO(n+1,1)$ --  
perhaps along the lines of Refs.~\cite{Guilloux:2016,Bhowmick:2018bmn}.

Just like for real Mellin amplitudes, the Mellin variable dependence in $p$-adic Mellin amplitudes enters solely via the Mandelstam-like invariants associated with internal lines.
In the Mellin-Barnes integral representation,  where the amplitude is expressed as a contour integral over lower-point contact amplitudes, these appear as arguments of local zeta functions, $\zeta_p$ and $\zeta_\infty$ in the $p$-adic and real cases respectively, and dictate the pole structure of the amplitude.
In both the real and $p$-adic cases, the complex contours in this representation  correspond to complex-shifting the internal dimensions of the bulk diagram. 
However, the complex manifold in the $p$-adic case is an infinite cylinder with the imaginary direction periodically identified, such that for each simple pole in the integrand in the $p$-adic case, the real analog features, in addition to the same pole, a semi-infinite sequence of poles corresponding to exchange of descendants. 
 
Consequently, due to the finite number of poles in the $p$-adic case, any Mellin amplitude is always expressible as a finite sum of ratios of elementary functions (precisely, the local zeta function $\zeta_p$), unlike the real case where closed-form expressions are typically not available and one must restrict to expressing the amplitudes in terms of increasingly intricate infinite sums or the Mellin-Barnes integral representation with unevaluated integrals~\cite{Fitzpatrick:2011ia,Paulos:2011ie}.

The careful reader may have noticed that the closed-form expressions for the $p$-adic Mellin amplitudes computed in this paper, given in \eno{contactFinal}, \eno{SingleFinal}, \eno{DoubleFinalA}, \eno{TripleLineAmp} and \eno{StarAmp}, appear to be hinting at a hidden structure obeyed by these amplitudes. 
A closer look at the expressions for the pre-amplitudes for each of these Mellin amplitudes also suggests that the pre-amplitudes themselves seem to be expressible in a structural form not very different from the full Mellin amplitudes.
These observations turn out to  be not mere coincidences, but can be formalized to reveal powerful recursion relations obeyed by the closed-form Mellin amplitudes as well as pre-amplitudes of {\it arbitrary} bulk diagrams at  tree-level~\cite{Jepsen:2018toappear}.

While in this paper we restricted our attention to $p$-adic Mellin amplitudes arising from bulk theories with polynomial couplings, amplitudes resulting from theories with derivative couplings may be readily extracted from the results obtained in this paper. 
This is because for a bulk action on the Bruhat--Tits tree, a polynomial coupling appears as a contact interaction vertex, while derivative couplings appear as nearest-neighbor interaction vertices.  
For this reason, any diagram constructed from derivative-couplings can be obtained from the sub-leading term of an exchange diagram in the limit where the internal operator is made infinitely heavy, see e.g.~Ref.~\cite{Gubser:2017tsi}.
Furthermore, it would be interesting to extract and interpret the flat-space limit~\cite{Polchinski:1999ry,Susskind:1998vk,DHoker:1999kzh,Gary:2009ae,Okuda:2010ym,Penedones:2010ue,Nandan:2011wc} of $p$-adic Mellin amplitudes, especially in light of the fact that not much is known about $p$-adic theories which could describe such flat-space  amplitudes.

We further restricted ourselves to only scalar fields in this paper. 
It would be interesting to relate and extend  the results of this paper to theories of particles with non-zero spin.
This has been a topic of much interest and recent progress in conventional AdS/CFT, see e.g.\ Refs.~\cite{Faller:2017hyt,Giombi:2017hpr,Sleight:2017fpc,Chen:2017xdz,Costa:2018mcg,Sleight:2018epi,Sleight:2018ryu}. 
On the $p$-adic front, however, it is at present not well understood how to describe spinning degrees of freedom in a discrete bulk geometry. A conceptual understanding of this is a  natural  next step worth pursuing.

Another promising avenue
is the study of $p$-adic Mellin amplitudes at loop level. 
Studying $p$-adic AdS/CFT at loop-level brings to fore the question of sub-AdS dynamics. Likely, a proper treatment should go beyond the discrete bulk tree geometry which was sufficient for our purposes  here. 
In fact in this paper, in the explicit calculation of Mellin amplitudes we passed to a continuum $p{\rm AdS}_{n+1}$ space~\cite{Gubser:2016guj} (see also Ref.~\cite{Bhowmick:2018bmn} for a related continuum construction), which is a refinement of the Bruhat--Tits tree, but purely for computational convenience since we restricted ourselves  to  a bulk-to-bulk propagator defined on the course-grained Bruhat--Tits tree. 
A natural generalization of the bulk-to-bulk propagator sensitive to sub-AdS length scales would possibly involve the chordal distance function of Ref.~\cite{Gubser:2016guj}. 
Indeed some work on constructing such an object recently appeared in Ref.~\cite{Qu:2018ned}, and provided evidence for non-trivial contributions to position-space loop amplitudes from small scales. It would be interesting to investigate this line of direction from the point of view of the formalism presented in this paper.

\section*{Acknowledgments}

C.\ B.\ J.\ and S.\ P.\ thank Steven S.\ Gubser, Matilde Marcolli, and Brian Trundy for useful discussions and encouragement. 
The work of C.\ B.\ J. was supported in part by the Department of Energy under Grant No. DE-FG02-91ER40671, by the US NSF under Grant No. PHY-1620059, and by the Simons Foundation, Grant 511167 (SSG).

\appendix

\section{Barnes Lemmas: Real and $p$-adic}
\label{BarnesSection}
As part of the motivation for why it was natural to have Mellin variables living on a complex cylindrical manifold, we mentioned  in section \ref{MELLINSPACE} that the Barnes lemmas~\cite{Barnes1,Barnes2} find close $p$-adic analogues in terms of contour integrals on the ``complex cylinder'' (see section \ref{MELLINSPACE} for a description of the complex cylinder). The analogy is most striking when these lemmas are re-expressed in terms of local zeta functions \eno{zetap} and \eno{zetainfty}, as we present below.
\vspace{1.5em}

\noindent \textit{The first Barnes lemma.}
\eqn{}{
\int_{-i\infty}^{i\infty}\frac{dz}{2\pi i}\,\zeta_\infty(a+z)\zeta_\infty(b+z)\zeta_\infty(c-z)\zeta_\infty(d-z) &=2\,\frac{\zeta_\infty(a+c)\zeta_\infty(a+d)\zeta_\infty(b+c)\zeta_\infty(b+d)}{\zeta_\infty(a+b+c+d)} \cr \cr 
\int_{-\frac{i\pi}{\log p}}^{\frac{i\pi}{\log p}}\frac{dz}{2\pi i}\,\zeta_p(a+z)\zeta_p(b+z)\zeta_p(c-z)\zeta_p(d-z) &= \frac{1}{\log p}\frac{\zeta_p(a+c)\zeta_p(a+d)\zeta_p(b+c)\zeta_p(b+d)}{\zeta_p(a+b+c+d)}\,.
}
The two above equations hold true when $a$, $b$, $c$, and $d$ are positive numbers so that the poles at $z=-a$ and $z=-b$ lie to the left of the contour and the poles at $z=c$ and $z=d$ lie to the right. 

\vspace{1.5em}
\noindent \textit{The second Barnes lemma.}
\eqn{}{
& \int_{-i\infty-|\epsilon|}^{i\infty-|\epsilon|}\frac{dz}{2\pi i}
\frac{\zeta_\infty(a+z)\zeta_\infty(b+z)\zeta_\infty(c+z)\zeta_\infty(d-z)\zeta_\infty(-z)}{\zeta_\infty(a+b+c+d+z)}
\cr
&\,= 2\,\frac{\zeta_\infty(a)\zeta_\infty(b)\zeta_\infty(c)\zeta_\infty(a+d)\zeta_\infty(b+d)\zeta_\infty(c+d)}{\zeta_\infty(b+c+d)\zeta_\infty(a+c+d)\zeta_\infty(a+b+d)} \cr
\cr 
& \int_{-\frac{i\pi}{\log p}-|\epsilon|}^{\frac{i\pi}{\log p}-|\epsilon|}\frac{dz}{2\pi i}
\frac{\zeta_p(a+z)\zeta_p(b+z)\zeta_p(c+z)\zeta_p(d-z)\zeta_p(-z)}{\zeta_p(a+b+c+d+z)}
\cr
&\,= \frac{1}{\log p}\,\frac{\zeta_p(a)\zeta_p(b)\zeta_p(c)\zeta_p(a+d)\zeta_p(b+d)\zeta_p(c+d)}{\zeta_p(b+c+d)\zeta_p(a+c+d)\zeta_p(a+b+d)}\,.
}
The above two equations hold true when $a$, $b$, $c$, and $d$ are positive numbers so that the poles at $z=-a$, $z=-b$, and $z=-c$ lie to the left of the contour while the poles at $z=0$ and $z=d$ on lie the right. $\epsilon$ is any non-zero real number such that $|\epsilon|$ is less than $a$, $b$, $c$, and $d$.

The $p$-adic versions of the Barnes lemmas presented above can be straightforwardly verified by an application of Cauchy's theorem by closing the contours to the left and summing over the enclosed residues.

\bibliographystyle{ssg}
\bibliography{mellin}

\end{document}